\documentclass[lettersize,journal]{IEEEtran}
\usepackage[caption=false,font=normalsize,labelfont=sf,textfont=sf]{subfig}
\usepackage[colorlinks,
            linkcolor=red,
            anchorcolor=red,
            citecolor=green]{hyperref}
\usepackage{tikz}
\usepackage{svg}
\usepackage{url}
\usepackage{multirow}
\usepackage{tabularx}
\usepackage[flushleft]{threeparttable}
\usepackage{enumitem}
\usepackage{adjustbox}
\usepackage{comment}
\usepackage[square, comma, sort&compress, numbers]{natbib}
\let\cite\citep
\setcitestyle{comma}
\setcitestyle{open={[},close={]},sep={,}}
\usepackage{graphicx}
\graphicspath{{figures/}}
\usepackage{amsmath,amssymb,amsfonts}
\usepackage{array}
\usepackage{xspace}  
\usepackage{algpseudocode}
\usepackage[linesnumbered,ruled,vlined]{algorithm2e}

\usepackage{multirow}
\usepackage{colortbl} 
\usepackage{hhline} 
\usepackage{array}

\usepackage{xcolor}
\usepackage{pifont}

\definecolor{lightgray}{gray}{0.9}
\definecolor{sotacolor}{RGB}{152,52,48}
\definecolor{runupcolor}{RGB}{236,84,40}

\newcommand{\laundry}{Laundry\xspace}
\newcommand{\slowrelease}{Slow Release\xspace}
\newcommand{\targetmapping}{Target Mapping\xspace}
\newcommand{\ourdefense}{TED-LaST\xspace}

\makeatletter
\DeclareRobustCommand\onedot{\futurelet\@let@token\@onedot}
\def\@onedot{\ifx\@let@token.\else.\null\fi\xspace}

\usepackage{multirow} 
\usepackage{textcomp}
\usepackage{stfloats}
\usepackage{verbatim}
\usepackage{blindtext}
\usepackage{fancyhdr}
\usepackage{booktabs}
\usepackage{siunitx}
\usepackage{fontawesome}
\usepackage{tablefootnote}
\usepackage{cuted}
\usepackage{capt-of}
\usepackage{wrapfig}
\usepackage{csquotes}

\usepackage[ruled,vlined]{algorithm2e}

\definecolor{color1}{HTML}{F39DA0} 
\definecolor{color2}{HTML}{95BCE5} 
\definecolor{color3}{HTML}{E84445} 
\definecolor{color4}{HTML}{1999B2} 
\MakeOuterQuote{"}

\begin{document}

\title{TED-LaST: Towards Robust Backdoor Defense Against Adaptive Attacks}

\author{
    \IEEEauthorblockN{Xiaoxing Mo\textsuperscript{1,*}, 
                      Yuxuan Cheng\textsuperscript{2}, 
                      Nan Sun\textsuperscript{3}, 
                      Leo Yu Zhang\textsuperscript{4}, 
                      Wei Luo\textsuperscript{1}, 
                      Shang Gao\textsuperscript{1}}\\
    \IEEEauthorblockA{\textsuperscript{1}Deakin University, Australia}\\
    \IEEEauthorblockA{\textsuperscript{2}Beijing Normal-Hong Kong Baptist University, China}\\
    \IEEEauthorblockA{\textsuperscript{3}University of New South Wales, Australia}\\
    \IEEEauthorblockA{\textsuperscript{4}Griffith University, Australia}
    \thanks{* Corresponding author: moxi@deakin.edu.au}
}

\maketitle



\begin{abstract}
Deep Neural Networks (DNNs) are vulnerable to backdoor attacks, where attackers implant hidden triggers during training to maliciously control model behavior.
Topological Evolution Dynamics (TED) has recently emerged as a powerful tool for detecting backdoor attacks in DNNs. However, TED can be vulnerable to backdoor attacks that adaptively distort topological representation distributions across network layers. 
To address this limitation, we propose \ourdefense (Topological Evolution Dynamics against \underline{La}undry, \underline{S}low release, and \underline{T}arget mapping attack strategies), a novel defense strategy that enhances TED's robustness against adaptive attacks. \ourdefense introduces two key innovations: label-supervised dynamics tracking and adaptive layer emphasis. These enhancements enable the identification of stealthy threats that evade traditional TED-based defenses, even in cases of inseparability in topological space and subtle topological perturbations.
We review and classify data poisoning tricks in state-of-the-art adaptive attacks and propose \textit{enhanced adaptive attack with target mapping}, which can dynamically shift malicious tasks and fully leverage the stealthiness that adaptive attacks possess.
Our comprehensive experiments on multiple datasets (CIFAR-10, GTSRB, and ImageNet100) and model architectures (ResNet20, ResNet101) show that \ourdefense effectively counteracts sophisticated backdoors like Adap-Blend, Adapt-Patch, and the proposed enhanced adaptive attack. 
\ourdefense sets a new benchmark for robust backdoor detection, substantially enhancing DNN security against evolving threats.
\end{abstract}


\section{Introduction}
\IEEEPARstart{D}{eep} Neural Networks (DNN) models have revolutionized fields such as computer vision~\cite{9127813}, speech recognition~\cite{8585066}, and autonomous driving~\cite{9420291} with their impressive capabilities. Despite these advances, their dependence on expansive datasets and complex training procedures introduces significant vulnerabilities, notably through backdoor attacks. Backdoor attacks implant hidden behaviors in DNN models, which can be activated by specific triggers. Remarkably, these backdoors do not impair the model's performance on clean data, making them particularly stealthy and damaging.

In classification tasks, these attacks typically involve poisoning the training dataset, where only a minor fraction of the training data is manipulated with attacker designated triggers. Once the model learns these triggers, it associates them with specific, attacker-defined classes. 
The development of backdoor attacks has evolved considerably, as evidenced by the seminal work of BadNets \cite{gu2017badnets} and subsequent developments \cite{li2021backdoor,liu2017trojaning,yang2021careful,adi2018turning,lin2020composite,tang2021demon,chen2017targeted,zeng2021rethinking,quiring2020backdooring,nguyen2020input,li2021invisible,duan2024conditional,10191661,10122715,9448491}, which vary in the intricacies of data poisoning timing and strategies.

Given the stealthy nature and potential harm of backdoor attacks, developing robust backdoor detection approaches has become paramount.
Backdoor defenses are typically categorized into three main groups based on their target of analysis: model-level \cite{xu2021detecting}, label-level \cite{liu2019abs,wang2019neural,guo2023universal,zhu2024neuralsanitizer}, and sample-level \cite{tran2018spectral,gao2019strip,tang2021demon,liu2023detecting,10646679}. Among these, sample-level defenses offer the most granular detection by identifying individual malicious samples as anomalies. These defensive approaches are effective particularly due to the key observations: a backdoored model often learns an excessively strong signal for the trigger within the latent space~\cite{qi2022revisiting}, overshadowing other semantic features and facilitating the clear separation of poisoned samples from clean ones. 

However, the separability between malicious inputs and normal inputs is not inherently guaranteed. Defenses can fail when the separability of malicious input representations from normal inputs in the latent space is deliberately suppressed \cite{qi2022revisiting}. This vulnerability of existing backdoor detection methods prompts attackers to devise \textbf{adaptive attacks} by either modifying the training process \cite{9230390,bagdasaryan2021blind} or implementing a bag of data poisoning tricks \cite{10646679, tang2021demon}. Furthermore, due to the versatility and broader applicability of data poisoning, its potential harm is significantly greater.
Some of the most commonly used tricks in this bag include \textbf{\laundry} \cite{qi2022revisiting,tang2021demon,lin2020composite,10646679,10191661,peng2024model}, which incorporates triggered samples but with correct labeling, \textbf{\slowrelease} \cite{xie2020dba,xue2020one}, which uses part of the trigger during training while retaining it intact during inference, and \textbf{\targetmapping}, which incorporates a shared trigger but targets diverse classes \cite{xiao2022multitarget, xue2020one}. 
Notably, for adaptive attacks, these data poisoning tricks are data-agnostic: the same trick can be applied to various types of poisoning data, and multiple tricks can be combined within one data poisoning attack.

As backdoor attacks become increasingly adaptive, defenders are continuously enhancing their detection capabilities. For example, to counter specific single tricks, defense methods have been developed particularly for \laundry~\cite{tang2021demon, 10646679} and for \slowrelease~\cite{hayase2021spectre}. Among these, our prior study demonstrated the effectiveness of using topological evolution dynamics (TED) to detect backdoor attacks with \laundry at the input level \cite{10646679}. TED analyzes the evolution of the topological representation of input samples as they propagate through the network, leveraging the observation that poisoned and clean samples often exhibit distinct evolutionary behaviors in topological space.

When facing combined tricks, such as \laundry and \slowrelease used in conjunction within one poisoning attack, this leads to the latent inseparability between malicious sample and clean sample in the metric space \cite{gao2019strip, hayase2021spectre, tang2021demon, liu2023detecting} or in the topological space.
Upon closer examination of this separability problem, we observe that the global topology feature utilized by TED cannot provide sufficient clarity. Furthermore, equal weighting to all layers becomes ineffective for malicious samples that propagate through multiple layers with subtle perturbations that resemble clean samples. 
Based on these observations, we propose two key insights:
(1)  \textbf{Malicious samples typically traverse longer trajectories} from their original class to the target class compared to samples from the target class itself. This extended trajectory in the feature space may provide a distinguishing characteristic for detection. (2) \textbf{Not all layers are equal}. The topological representation differences between malicious and clean samples may vary across layers. Therefore, we need to dynamically identify key layers and assign different weights to different layers during outlier detection.
These insights motivate us to develop a more fine-grained and robust approach for detecting adaptive attack sample or even subtle-perturbation samples in the topological space.

In response, we present \textbf{\ourdefense{}} (Topological Evolution Dynamics against \underline{La}undry, \underline{S}low Release, and \underline{T}arget Mapping attack strategies), a novel topology-based backdoor detector extending our previous work \cite{10646679} for robustness against adaptive attacks. \ourdefense{} leverages supervised label information and modularity-based adaptive layer emphasis to improve detection robustness and detect malicious samples with subtle-perturbations in extreme cases, even when the topological separability between benign and malicious samples has been severely compromised.
Our key contributions can be summarized as follows:

\begin{itemize}

\item This study carefully reviews and classifies data poisoning tricks in SOTA adaptive backdoor attacks (Section~\ref{sec:Backdoor in Deep Neural Network}), revealing the drawbacks of existing backdoor detectors that aim to separate benign and malicious samples in the metric space (Section~\ref{sec:Backdoor in Deep Neural Network}) or topological space (Section~\ref{sec:case study of ted against adaptive attack}). We show that adaptive attacks and our proposed Enhanced Adaptive Attacks can invalidate SOTA detectors by obscuring sample features.

\item This study proposes \ourdefense, which significantly enhances the robustness of topology-based backdoor detectors against adaptive attacks by quantifying malicious sample perturbation to address the inseparability between malicious and benign samples in topological space and prioritizing informative topological features to address insensitivity to subtle perturbations (Section~\ref{sec:Adaptive TED Design and Performance Against Adaptive Attacks}).

\item This study extensively validates \ourdefense{} across various scenarios, demonstrating that  it achieves precision higher than 90\% and F1 score higher than 85\% against all SOTA adaptive attacks and Enhanced Adaptive Attacks (Section~\ref{sec:Adaptive TED Design and Performance Against Adaptive Attacks}). Our results consistently show that \ourdefense{} outperforms SOTA defenses.

\end{itemize}

\section{Backdoor in Deep Neural Networks}
\label{sec:Backdoor in Deep Neural Network}
A DNN model, denoted as \( f \), comprises a sequence of layers \(\{ f_l : l \in [1, N] \}\), where each layer functions as a transformation. For an input \( x \), the output of the neural network \( f \) is computed by the composition:
\begin{IEEEeqnarray}{rCl}
f(x) = \big(f_N \circ \cdots \circ f_1\big)(x).
\end{IEEEeqnarray}
Following previous studies~\cite{gu2017badnets,bagdasaryan2021blind,gao2019strip,lin2020composite,tang2021demon, 10646679, qi2022revisiting}, this paper focuses on DNN models applied to classification tasks.
Specifically, we address a classification problem where the input space is denoted as \( \mathcal{X} \) and the set of all classes as \( \mathcal{Y} \). Each ground-truth input-output pair $(x, y)$ is a sample, where $x \in \mathcal{X}$ and $y \in \mathcal{Y}$.
The training dataset, denoted as \( \mathcal{D} = \{(x_i, y_i)\} \), consists of data points \((x_i, y_i)\) and the model \( f^* \) is trained to minimize a loss function \( L(\cdot, \cdot) \) over $\mathcal{D}$ as:
\begin{IEEEeqnarray}{rCl}
f^* = \arg\min_f \sum_i L(y_i, f(x_i)). \label{eq:standard_classification}
\end{IEEEeqnarray}

\subsection{Backdoor Attacks}
Backdoor attacks embed malicious functionalities into neural networks, causing abnormal behavior only for trigger-carrying inputs.  While some attacks directly modify model parameters \cite{dumford2020backdooring} or the model structure \cite{tang2020embarrassingly}, the most common 
approach involves dirty-label data poisoning. This typically involves embedding triggers into a subset of training samples from a source label and changing their labels to the target label. The work in \cite{gu2017badnets} first demonstrated this technique, showing that stamping an image with a small fixed pattern (\textit{e.g.}, a white square) can successfully create a backdoor.
These altered samples, denoted as $A(x)$ for an original input $x$, are labeled with the attacker's chosen target class $c_\text{target}$, creating a poisoned dataset $\mathcal{D}_\text{p} = \{(A(x), c_\text{target}) \mid (x, y) \in \mathcal{D}\}$, where $\mathcal{D}$ is the original, clean dataset. When trained on the combined dataset $\mathcal{D} \cup \mathcal{D}_\text{p}$, the model $f$ learns to classify triggered samples $A(x)$ as the target $c_\text{target}$ while still maintaining high accuracy on clean samples from $\mathcal{D}$ \cite{gu2017badnets,chen2017targeted,liu2017trojaning,zeng2021rethinking}.

One common method for increasing the stealthiness of backdoor attack is to modify the training process itself, in addition to poisoning the data by embedding triggers in the samples. 
Examples include using generative networks to create dynamic triggers \cite{nguyen2020input, li2021invisible}, and incorporating specialized loss functions \cite{bagdasaryan2021blind}. 
However, attackers also seek simple yet effective methods, even under the constraint where only samples and associated labels can be modified without altering the training process, to effectively evade defenses.

\begin{figure}[t]
\centering
\hspace*{-0.4cm}
\begin{tabular}{@{}c@{\hspace{4mm}}c@{}}
\textbf{Adap-Blend} \cite{qi2022revisiting} & \textbf{Adap-Patch} \cite{qi2022revisiting} \\[1ex]
\begin{tabular}{@{}c@{}c@{}c@{}}
\includegraphics[width=0.22\linewidth, angle=90]{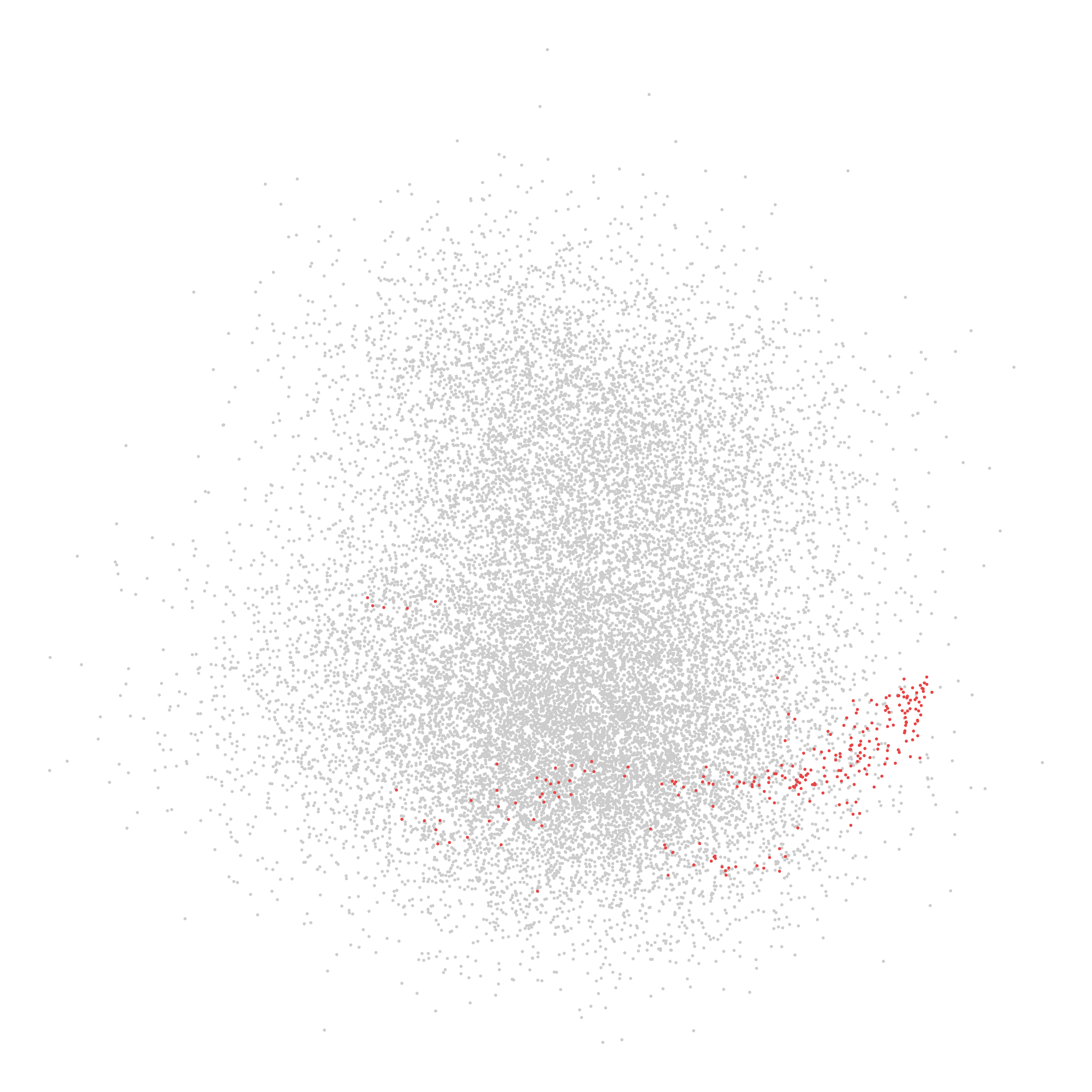} &
\hspace{-1mm}\tikz{\draw[gray,dashed,line width=0.5pt] (0,-6ex) -- (0,6ex);}\hspace{-1mm} &
\includegraphics[width=0.22\linewidth]{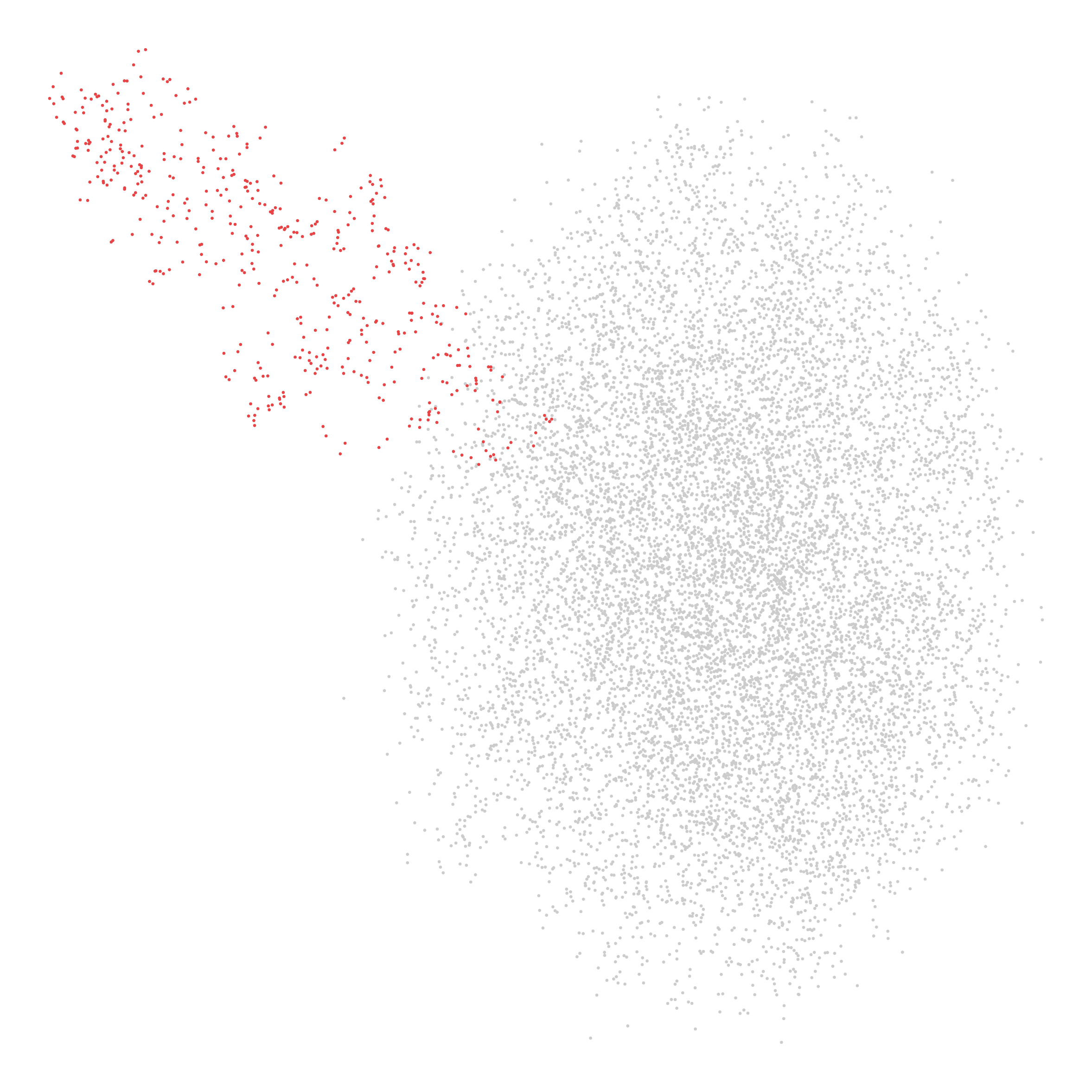} \\
\small{TED} & & 
\small\ourdefense{} (\textbf{Ours})
\end{tabular} &
\begin{tabular}{@{}c@{}c@{}c@{}}
\includegraphics[width=0.22\linewidth]{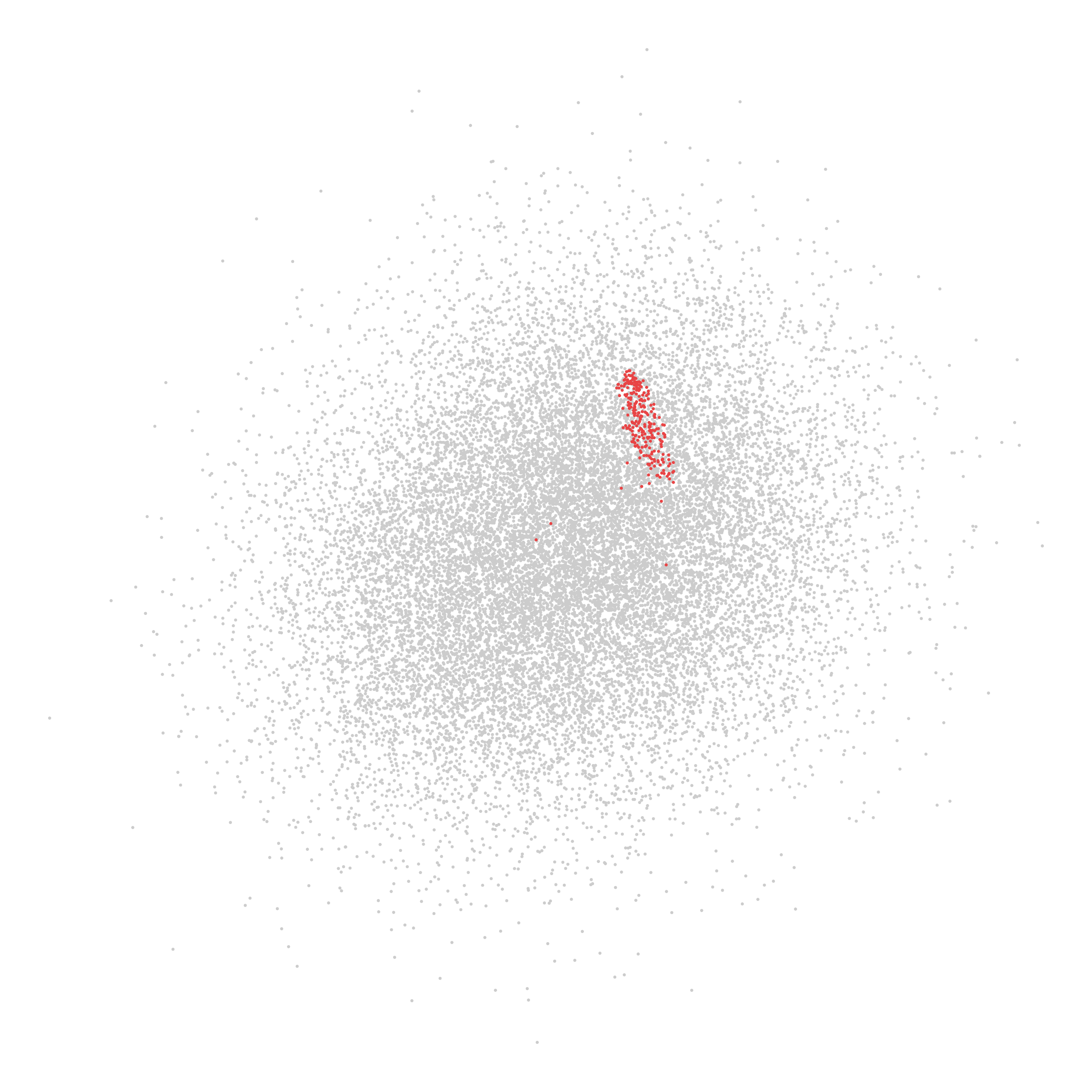} &
\hspace{-1mm}\tikz{\draw[gray,dashed,line width=0.5pt] (0,-6ex) -- (0,6ex);}\hspace{-1mm} &
\includegraphics[width=0.22\linewidth, angle=90]{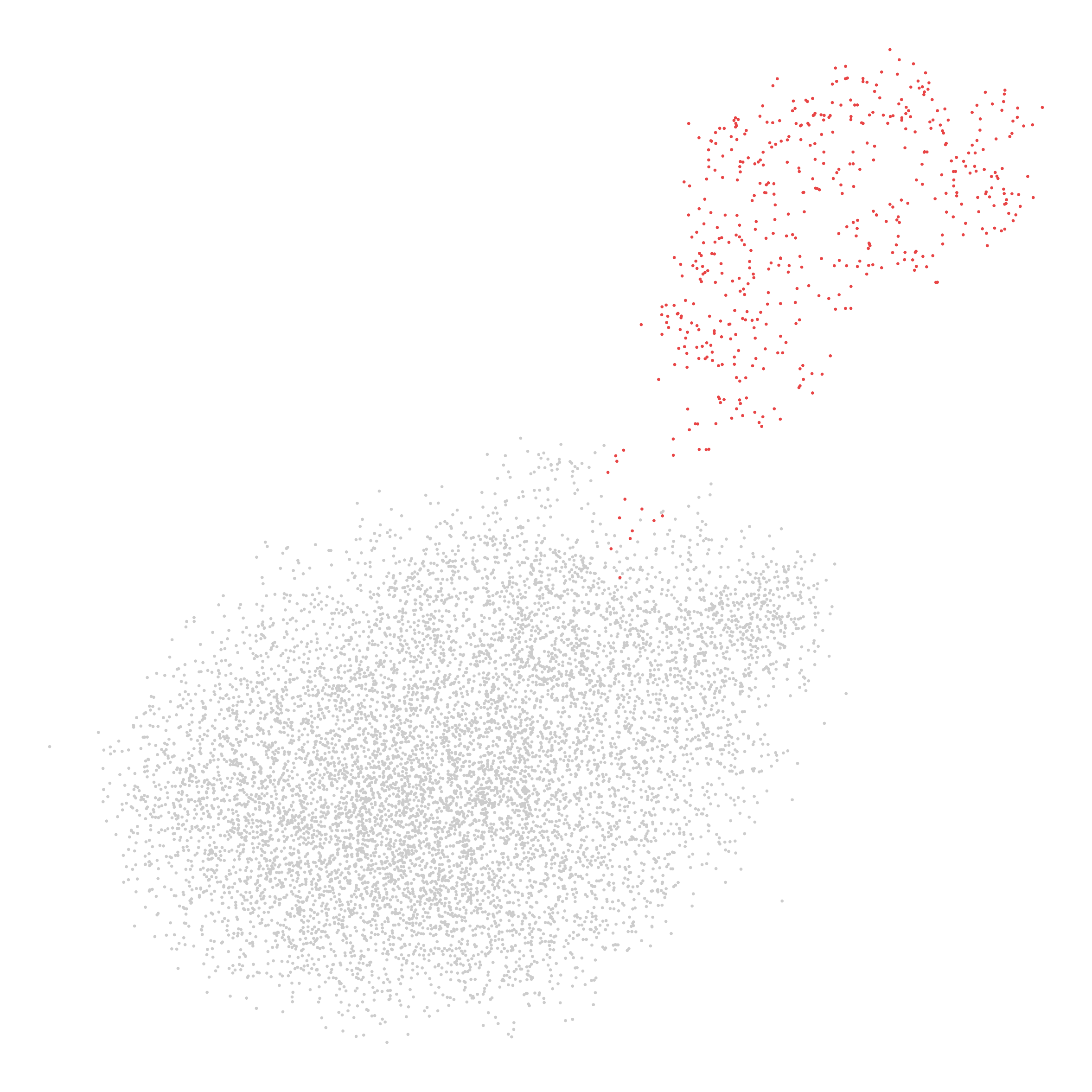} \\
\small{TED} & & 
\small{\ourdefense{}} (\textbf{Ours})
\end{tabular}
\end{tabular}
\caption{T-SNE visualization of sample feature vector separability for TED and \ourdefense{} on backdoored CIFAR-10 models under Adap-Blend (left) and Adap-Patch (right) attacks. For each attack, the left subplot shows results for TED, while the right subplot shows results for \ourdefense{}. \textcolor{red}{Red} points indicate malicious samples and \textcolor{gray}{Gray} points represent clean samples. The plots demonstrate improved separation between malicious and clean samples achieved by \ourdefense{} compared to TED.}
\label{fig:tsne_plots}
\end{figure}

\subsection{Bag of Data Poisoning Tricks for Adaptive Attack}
\label{sec:Adaptive Attack}

\subsubsection{\laundry}
\laundry, firstly studied in \cite{tang2021demon}, is a trick that incorporates training samples with the trigger while retaining their correct labels, allowing attacks to evade defenses~\cite{peng2024model,10646679}. This approach prevents the backdoored model from learning an overwhelmingly strong signal (more detectable in defense) for the trigger that would always lead to the target class~\cite{qi2022revisiting}. 
During training, two types of triggered samples are used: 1) poisoned samples with triggers labeled as the target class \(c_\text{target}\), and 2) samples with triggers that maintain their original labels. This trick can be formulated as:
\begin{IEEEeqnarray}{rCl} 
\mathcal{D}_\text{p} &=& \{ (A(x),  c_\text{target}) \mid   (x,y) \in \mathcal{D}  \},  \label{Eq:backdoor dataset} \\ 
\mathcal{D}_\text{l} &=& \{(A(x), y) \mid  (x,y) \in \mathcal{D}  \},  \label{Eq:laundry dataset}
\end{IEEEeqnarray}
where $\mathcal{D}_\text{p}$ represents the poisoned dataset and $\mathcal{D}_\text{l}$ denotes the \laundry dataset. Both datasets collectively constitute the model's training set.

\subsubsection{Slow Release}
\slowrelease, firstly studied in \cite{xie2020dba}, is a trick that incorporates training samples with partial trigger during training, while using the complete trigger during inference to activate the backdoor \cite{xie2020dba,xue2020one}. This approach gradually introduces the backdoor, weakening the strong signal for the trigger learned by the model. The diverse partial triggers during training prevent the model creates a sudden, easily detectable correlation between the trigger and the target class \cite{hayase2021spectre}. 

For instance, during training, only portions of the trigger pattern are applied (\textit{e.g.}, using only two patches of a four-patch trigger), while the complete trigger pattern is restored during inference/attack phase. More formally, during training, we use a set of parameters $\mathcal{R}_\text{t}$ (a subset of full parameter space $\mathcal{R}$) to control the trigger intensity or geometric attribute $\beta_\text{t}$, while during inference, we use a mapping function $g(\beta_\text{t})$ to convert the training-phase parameter to its full-strength inference counterpart. The poisoned dataset can be formulated as:
\begin{IEEEeqnarray}{l}
\!\!\mathcal{D}_\text{p} = \big\{\big(A_\text{trig}(x, \beta_\text{t}),  c_\text{target}\big) \mid (x,y) \in \mathcal{D}, \beta_\text{t} \in \mathcal{R}_\text{t} \subset \mathcal{R}\big\},   \label{Eq:slowreleasedataset}\\
\!\!A_\text{trig}(x, \beta) = x \oplus I(\delta, \beta),  \label{Eq:amod}
\end{IEEEeqnarray}
where $A_\text{trig}$ is the trigger application function that combines the trigger with the input sample, $I(\delta, \beta)$ is a function that generates the trigger pattern $\delta$ based on a parameter $\beta$, which controls either the trigger's intensity (\textit{e.g.}, opacity) or geometric attributes (\textit{e.g.}, size, gap, location).

\begin{table}[h]
\centering
\caption{Backdoor Attack Notation.}
\label{tab:terms}
\begin{tabularx}{\columnwidth}{lX}
\toprule
Term & Description \\
\midrule
$T(\cdot)$ & \targetmapping function \\
$A(x)$ & Function creating new sample from $x$ \\
$A_\text{trig}(x, \beta)$ & Function modifying $x$ with trigger controlled by $\beta$ \\
$A_\text{trig}(x)$ & Function applying trigger for Source-Specific \targetmapping \\
$c_\text{target}$ & Attacker's chosen target class \\
$\mathcal{D}_\text{p}$, $\mathcal{D}_\text{l}$ & Backdoor poisoning and \laundry datasets \\
$I(\delta, \beta)$ & Function modulating trigger $\delta$ with $\beta$ \\
$\mathcal{R}$ & Set of trigger attribute parameters, $\beta \in \mathcal{R}$ \\
$\mathcal{R}_\text{t}$ & Subset of $\mathcal{R}$ for \slowrelease training phase \\
$g(\beta)$ & Function mapping training to inference-time parameter \\
$\mathcal{S}$ & Set of source classes \\
$\oplus$ & Operation of mixing trigger with input \\
\bottomrule
\end{tabularx}
\end{table}

\subsubsection{\targetmapping}
\label{sec:targetmapping}

\targetmapping, firstly studied in \cite{xue2020one}, is a data poisoning trick that use a single, shared trigger to create diverse mappings to multiple target classes~\cite{edraki2021odyssey,xiao2022multitarget, xue2020one,bagdasaryan2021blind}.
Instead of creating a one-to-one relationship (\textit{e.g.}, one trigger to target class A), \targetmapping establishes a one-to-many backdoor mapping (\textit{e.g.}, one trigger to target classes A,  B, and C). The backdoor's malicious behaviors thus become dependent on factors \textbf{beyond} just the trigger. While the trigger acts as one element, other seemingly benign features in the input data (such as specific pixel combinations or values within a particular range) can dictate which malicious task the model activates.
The target mapping function is defined as:
\begin{IEEEeqnarray}{rCl}
{T: (\mathcal{S} \cup \{\emptyset\}) \times (\mathcal{R} \cup \{\emptyset\}) \rightarrow \mathcal{Y}},
\label{eq:targetmapping_function}
\end{IEEEeqnarray}
where $\mathcal{S}$ denotes the set of source classes.
For scenarios focusing solely on the source class, known as Source-Specific (SS) \targetmapping, the poisoning dataset is:
\begin{IEEEeqnarray}{rCl}
\mathcal{D}_\text{p} &=& \big\{(A_\text{trig}(x), T(y)) \mid (x,y) \in \mathcal{D}, y \in \mathcal{S} \big\},
\label{eq:ss_targetmapping}
\end{IEEEeqnarray}
where \(A_\text{trig}(x) = x \oplus \delta\). SS applies the trigger uniformly without differentiating trigger attribute \(\beta\).
Unlike SS, the Source-Specific \& Trigger Attribute (SS\&TA) \targetmapping considers both source class and trigger attributes:
\begin{IEEEeqnarray}{rCl}
\mathcal{D}_\text{p} &=& \big\{\big(A_\text{trig}(x, \beta), T(y, \beta)\big) \mid (x,y) \in \mathcal{D}, \big. \nonumber \\
&& \big. y \in \mathcal{S} \cup \{\emptyset\}, \beta \in \mathcal{R} \cup \{\emptyset\}\big\}.
\label{eq:ssta_targetmapping}
\end{IEEEeqnarray}

\subsubsection{Combining Tricks}

As backdoor detection methods advance, single-trick attacks have become increasingly vulnerable to detection. In response, attackers have developed more adaptive attacks such as Adap-Blend and Adap-Patch, which combine \laundry and \slowrelease \cite{qi2022revisiting}.
The Adap-Blend attack uses partitioned, low-opacity triggers for training, and full, higher-opacity triggers for attacks. The Adap-Patch attack utilizes multiple small and diverse patches as triggers, employing combinations of fully opaque patches during the attack.
To implement \laundry, both Adap-Blend and Adap-Patch inject triggers into a portion of clean samples during training, while keeping their true labels unchanged. Such adaptive attacks not only evade SOTA detection methods that rely on latent feature separability in metric space but also circumvent TED, which utilizes feature separability in topological space, as illustrated in Fig.~\ref{fig:tsne_plots}.

\subsubsection{Enhanced Adaptive Attack}
Building upon these advanced combinations, we propose to further integrate even more data-agnostic tricks from the bag into the adaptive attack framework. Particularly, models compromised by a \targetmapping attack can alternate between multiple malicious tasks based on the presence of attacker-chosen backdoor features. Importantly, this task-switching capability persists even when attacks \textbf{share} a common trigger pattern \cite{lin2020composite,xue2020one,xiao2022multitarget,bagdasaryan2021blind}, meaning the same trigger no longer consistently leads to a static malicious outcome.

Unlike Adap-Blend and Adap-Patch which are implemented with a static target class, we introduce \textbf{Enhanced Adaptive Attacks}, which combine \laundry(L), \slowrelease(SR), and \targetmapping (SS or SS\&TA) to create dynamic-target adaptive attacks. The key idea is to modify the trigger according to Eq.~\eqref{Eq:amod}, thereby extending the poisoning datasets $\mathcal{D}_\text{p}$ (as defined in Eq.~\eqref{eq:ssta_targetmapping}) and \laundry dataset $\mathcal{D}_\text{l}$ (as defined in Eq.~\eqref{Eq:laundry dataset}). Specifically, to backdoor a model using the SS\&TA+L+SR attack, we train the model by minimizing a loss function comprised of three components: clean loss ($L_\text{c}$), laundry loss ($L_\text{l}$), and poison loss ($L_\text{p}$). Building upon Eq.~\eqref{eq:standard_classification}, our complete loss function is:
\begin{IEEEeqnarray}{lCl}
f^* &=& \arg\min_f (L_\text{c} + L_\text{l} + L_\text{p}), \label{eq:objective_function_main} \\
L_\text{c} &=& \sum_{(x_i, y_i) \in \mathcal{D}} L(y_i, f(x_i)), \label{eq:clean_loss} \\
L_\text{l} &=& \sum_{(x_i, y_i) \in \mathcal{D}_\text{l}} L(y_i, f(A_\text{trig}(x_i, \beta_{t,i}))), \label{eq:laundry_loss} \\
L_\text{p} &=& \sum_{(x_i, y_i) \in \mathcal{D}_\text{p}} L(T(y_i, g(\beta_{t,i})), f(A_\text{trig}(x_i, \beta_{t,i}))). \label{eq:poison_loss}
\end{IEEEeqnarray}
Additional configurations of Enhanced Adaptive Attacks, including SS+L+SR, are detailed in Appendix~\ref{appendix:Adaptive Attack}. 


\subsection{Existing Defenses Against Adaptive Attacks}
\label{Existing Defenses Against Adaptive Attacks}
Backdoor defenses are typically categorized into three main groups based on their analysis target: model-level, label-level, and sample-level.
Model-level defenses focus on analyzing the model itself.  For example, a meta-classifier can be trained on a set of clean and trojaned models to identify compromised models \cite{xu2021detecting}.
Label-level defenses aim to reverse-engineer potential triggers and remove inserted backdoors \cite{liu2019abs,wang2019neural,zhu2024neuralsanitizer}, or analyze anomalies in learned representations \cite{guo2023universal,chen2018detecting}. However, for model and label-level defenses, understanding why a model is flagged as compromised can be particularly challenging, especially under adaptive attacks that subtly alter model behavior. By contrast, sample-level defenses provide a more granular approach.

Sample-level defenses analyze the input data representations and model behavior.  For instance, SCAn \cite{tang2021demon} 
uses robust statistics to analyze the representation distribution across classes and employs a bi-component model to disentangle class identity and variations.  STRIP \cite{gao2019strip} detects triggers by overlaying input images on random samples and analyzing the entropy changes in the output labels.  TeCo \cite{liu2023detecting} assesses the model's corruption robustness, identifies distinct patterns of triggered samples under various image corruptions, and quantifies the consistency of model responses as corruption severity increases. 

Nevertheless, these sample-level defenses are not impervious to adaptive attacks. Specifically, adaptive attacks can diminish SCAn's effectiveness by suppressing the potential separation between clean and poisoned representations. While STRIP is effective against standard attacks, adaptive techniques can manipulate the entropy distribution of triggered inputs, thus blurring their distinction from benign data. Moreover, adaptive attacks can circumvent TeCo by engineering consistent behavior across corruption levels.  The detailed results of sample-level defenses against various adaptive attacks can be found in our experimental analysis in Section~\ref{sec:Design of Backdoor Attacks and Experimental Setup}.

\section{Original TED Encounters Adaptive Attack: A Case Study}
\label{sec:case study of ted against adaptive attack}

Topological Evolution Dynamics (TED) leverages the evolution of neural network activations in the topological space to outlier malicious samples. The TED feature vector is a measure that captures how a sample's activations align with its predicted class throughout the network.
specifically, TED feature vector quantifies the evolution of a sample's feature representation across different layers by tracking its relative position within the topological space of activations.
At each layer, it ranks the closeness of a sample's activations to those of its predicted class. By tracking these rankings across layers, TED maps each sample's evolutionary path through the network. 

TED adopts a topological approach to model feature spaces, focusing on relative proximity rather than mere vector distances. This involves defining a metric space $(\mathcal{V}, d)$, where $\mathcal{V}$ is a set of vectors or matrices, and $d$ is a metric function mapping any two elements in $\mathcal{V}$ to a non-negative real number. Each input $x$ at layer $l$ is represented as $h_l(x) = v \in \mathcal{V}^{(l)}$. In this space, an open ball centered at $v$ with radius $r$, denoted as $\mathcal{B}(v, r)$, includes all points $v'$ for which $d(v, v') < r$. These open balls form a topology based on neighborhood closeness.
For a benign sample $x_u$ with label $y_u$ at the layer $l$, there exists another sample $x'$, 
also labeled $y_u$, 
within a minimal radius \( r_l \)
such that $h_l(x')$ 
falls within $\mathcal{B}(v_u^{(l)}, r_l)$. This minimal radius \( r_l \) (1 by default) captures the local neighborhood structure around the sample $x_u$ at layer $l$.This assumption implies that benign samples of the same class will exhibit similar activation patterns within a certain proximity.

For each input sample, TED generates a ranking list at each network layer based on its proximity to other samples from the same predicted class. The TED feature vector for an input $x$ is defined as:
\begin{IEEEeqnarray}{rCl}
\text{TED}(x) = [K_1(x), K_2(x), \ldots, K_l(x), \ldots, K_N(x)].\label{eq:tedfeaturevector}
\end{IEEEeqnarray}
This sequence captures \( x \)'s topological evolution across the \( N \) network layers. Here, \( K_l(x) \) represents the rank of \( x \)'s nearest neighbor from its predicted class at layer \( l \), typically calculated using Euclidean distance. This sequential data reveals whether \( x \) consistently aligns with its class's typical activation patterns or diverges, potentially indicating an anomaly. An outlier detector is then trained on the TED feature vectors of benign samples from all classes. This detector flags inputs with significantly different TED trajectories as potential outliers (malicious).

\subsection{Limitations of TED}
\subsubsection{Inseparability in Topological Space Outlier Detection}
\label{subsection:TEDLimitationAgainstAdatptiveAttack}

Traditional attacks without adaptive approaches typically produce malicious data that exhibits significantly different characteristics from clean data in the topological space. Thus, TED assumes that malicious data has notable differences in topological evolution compared to clean data from all classes, and identifies malicious data through outlier detection.

However, this assumption becomes less robust against adaptive attacks. Such adaptive attacks manipulate the feature space and induce \textbf{inseparability} in the topological space of learned representations, as shown in Fig.~\ref{fig:tsne_plots}. This induced inseparability disrupts TED's ability to distinguish between benign and malicious samples.

The core vulnerability lies in TED's training procedure for its outlier detector. By using benign samples from all classes during training, the detector learns a broader notion of "normal" behavior, encompassing variations across all classes. As a result, adaptive attacks can craft malicious samples that fall within this broader range of acceptable normal variation, while still achieving their malicious objectives.

Such vulnerability is particularly susceptible to exploitation by adaptive attacks, since an adaptive attacker can construct malicious samples that are sufficiently close to any class, not necessarily the target class, to evade detection, rendering TED ineffective. This fundamental limitation of considering all classes equally during outlier detector training underscores the need for more focused approaches.

\begin{figure}[t!]
\centering
\includegraphics[width=0.7\linewidth]{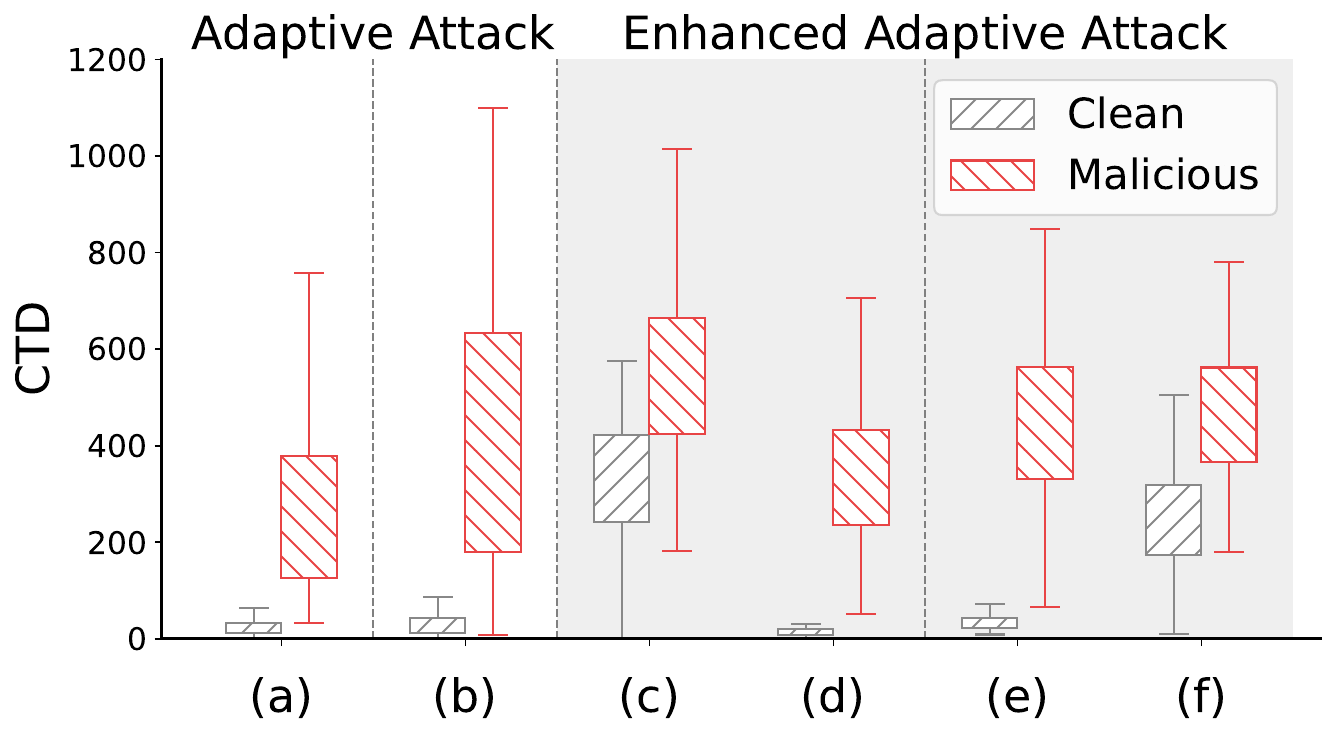} 
\caption{Cumulative Topological Distance (CTD) or different attack scenarios on CIFAR-10. (a) Adap-Blend, (b) Adap-Patch, (c) SS+L+SR Target Class A, (d) SS+L+SR Target Class B, (e) SS\&TA+L+SR Target Class A, (f) SS\&TA+L+SR Target Class B. Scenarios (a) and (b) represent adaptive attacks, while (c)-(f) represent Enhanced Adaptive Attacks.}
\label{fig:cdtplot}
\end{figure}

\begin{figure*}[t!]
\centering
\begin{tabular}{@{}c@{\hspace{1mm}}c@{}}
\includegraphics[width=0.48\linewidth, trim={0 0 0 130pt}, clip]{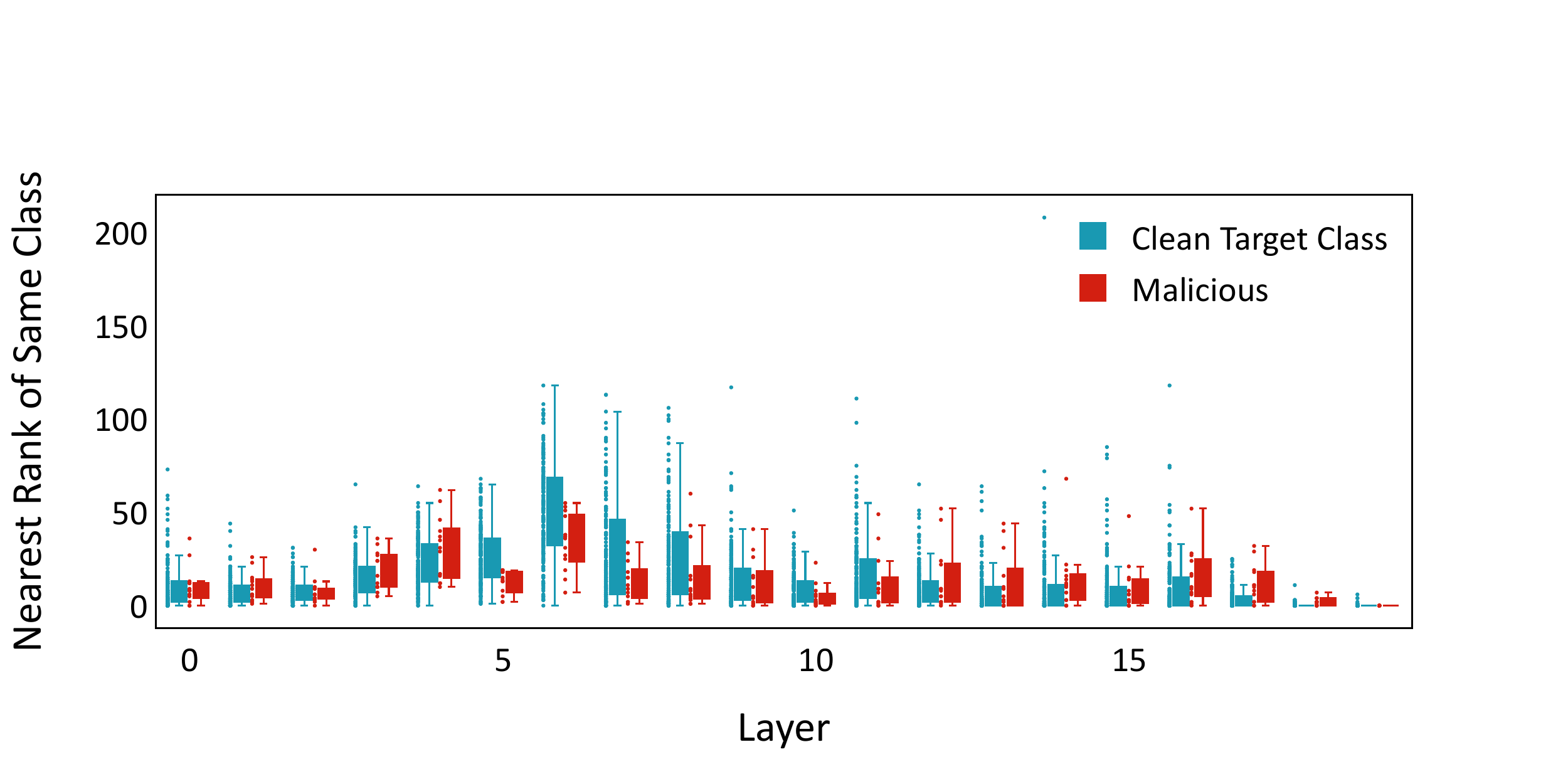} &
\includegraphics[width=0.48\linewidth, trim={0 0 0 130pt}, clip]{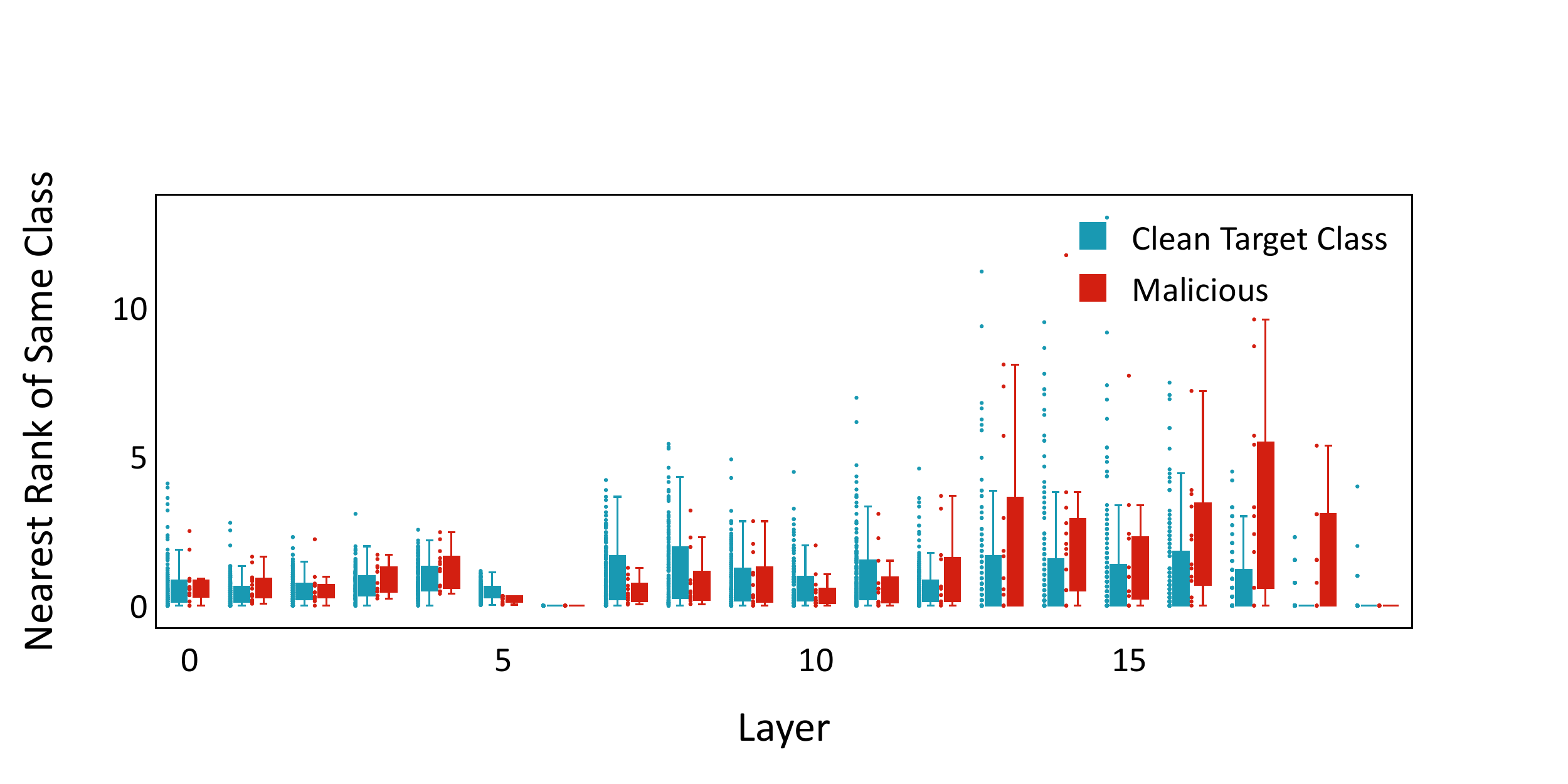} \\
(a) Non-weighted & (b) Modularity-based weighted
\end{tabular}
\caption{(a) Box plot of the topological feature vector on ResNet20 under Adap-Patch. The plot reveals CIFAR-10 malicious samples with subtle perturbations, which are too minor for class-wise TED to effectively identify as anomalies. (b) Box plot after applying modularity-based adaptive layer weighting, showing improved separation between clean and malicious samples.}
\label{fig:combined_visualization}
\end{figure*}

\subsubsection{Insensitive to Subtle Perturbations}
\label{subsec:Limitations of TED in Detecting Subtle Perturbations}

In adaptive attacks, malicious samples closely emulate the topological evolution of the target class, effectively "shadowing" legitimate samples' trajectories by maintaining minimal distance to their target class neighbors across multiple layers. Consequently, TED's feature vector exhibits limited resolution in distinguishing these samples.
The core challenge lies in TED's insufficient sensitivity to detect subtle yet persistent deviations from the true class trajectory. While TED effectively identifies significant topological shifts, it struggles to flag samples exhibiting only slight offsets in the topological space across layers. This limitation can lead to misclassification of samples whose TED feature vector falls within the margin of error for the target class. 

Such vulnerability is particularly susceptible to exploitation by adaptive attacks, which can engineer malicious samples to remain consistently within TED's detection limits across network layers by minimizing perturbations and targeting layers where TED's sensitivity is lower.

\subsection{Insight and Proposed Enhancements to TED}

\subsubsection{Label-Supervised Dynamics Tracking}
\label{subsection:Label-Supervised Dynamics Tracking}

The key insight is that malicious samples, originating from a different source class, traverse \textbf{a larger topological distance} compared to benign samples within the target class itself. To empirically quantify this difference in traversed distance, we introduce the Cumulative Topological Distance (CTD) metric:
\begin{IEEEeqnarray}{rCl}
\label{eq:ctd}
\text{CTD}(x) &=& \sum_{l=1}^{N-1} |K_{l+1}(x) - K_l(x)|, 
\end{IEEEeqnarray}
where $\mathcal{X}$ denotes the set of all samples. Here, $N$ is the total number of network layers, and $K_l(x)$ represents the ranking of sample $x$ at layer $l$.
As shown in Fig.~\ref{fig:cdtplot}, malicious samples consistently exhibit higher CTD values compared to clean samples predicted in the target class across all attacks. This observed discrepancy in the traversed distance suggests a potential distinction in the topological features of manipulated samples, motivating our label-supervised dynamics tracking approach for detecting them.

We then shift from a global to a class-specific perspective. We assume that the defender can access to a small set of clean samples with the correct label. 
Specifically, our method employs class-specific PCA-based outlier detection models.
The model computes a class-specific threshold using a reject parameter $\alpha$. This threshold is set to preserve the $(1 - \alpha)$ percentage of the variance explained by the principal components within the distribution of benign samples for that particular class. Samples exceeding this threshold are flagged as potential outliers.
Leveraging this label information, we establish a granular understanding of the expected data patterns within each class. This allows us to detect deviations that would otherwise be masked by the global topological blurring caused by adaptive attacks. By focusing on the unique characteristics of each class, we can identify malicious samples even when they blend seamlessly with their target class in the global topological space.

\subsubsection{Adaptive Layer Emphasis}

While incorporating label-supervised dynamics tracking mitigates the global inseparability issue of the original TED, detecting malicious samples that closely mimic the target class in the topological space remains challenging, especially for samples with subtle perturbations. These subtle perturbations result in lower CTD values for malicious samples, indicating smaller topological distances traversed across network layers. Consequently, their feature representations become more similar to those of benign samples in the target class, making them harder to detect as outliers. Empirical observations show that the distribution of ranks for malicious samples, particularly those with CTD values in the lower quartile of their distribution, shows considerable intersection with the distribution of ranks for benign samples in the target class across most layers in the TED (Fig.~\ref{fig:combined_visualization}a).

Previous studies have shown that assigning greater weight to key layers, typically the last few layers, can improve backdoor malicious detection \cite{huang2022backdoor, 10.1007/978-3-031-33377-4_33}. However, merely changing the number of layers does not effectively enhance TED performance \cite{10646679}. Therefore, in addition to identifying key layers at the network's end, we must also consider the inherent variability in clean samples from the front and middle layers. To address this, we propose a method to dynamically identify and emphasize key layers across the entire network. 

We adopt a modularity concept due to its effectiveness in quantifying cluster separation \cite{doi:10.1073/pnas.0601602103}. Our proposed modularity-based method quantifies the variability in each layer's feature space and adaptively adjusts the weights of different layers accordingly.
Specifically, modularity quantifies the degree to which a network can be divided into distinct communities or clusters \cite{Fortunato_2010}. We adapt this concept to the feature space, treating data points as nodes and their similarities as edge weights. The modularity score is computed by comparing the density of connections within clusters to the expected density if the connections were distributed randomly. A higher modularity score indicates a clearer separation between classes, implying the presence of well-defined clusters.

In our approach, layers with lower variability levels (higher modularity) are assigned greater weight in the final calculation of the feature vector, thereby emphasizing their contribution to the overall detection process. Conversely, layers with higher variability levels (lower modularity) are assigned smaller weight, thus reducing their contribution. By selectively emphasizing layers with different class distinctions, as depicted in Fig.~\ref{fig:combined_visualization}b, we enhance the method's resilience against the subtle differences introduced by adaptive attacks. In addition to addressing static target adaptive attacks, we provide a visualization comparing the non-weighted and modularity-based weighted methods under a dynamic-target Enhanced Adaptive Attack in Appendix~\ref{appendix:Subtle Perturbations on Noise Quantification and Adaptive Weighting}.


\section{Defense Design}
\label{sec:Adaptive TED Design and Performance Against Adaptive Attacks}

\begin{figure}[t]
\centering
\includegraphics[width=1\linewidth]{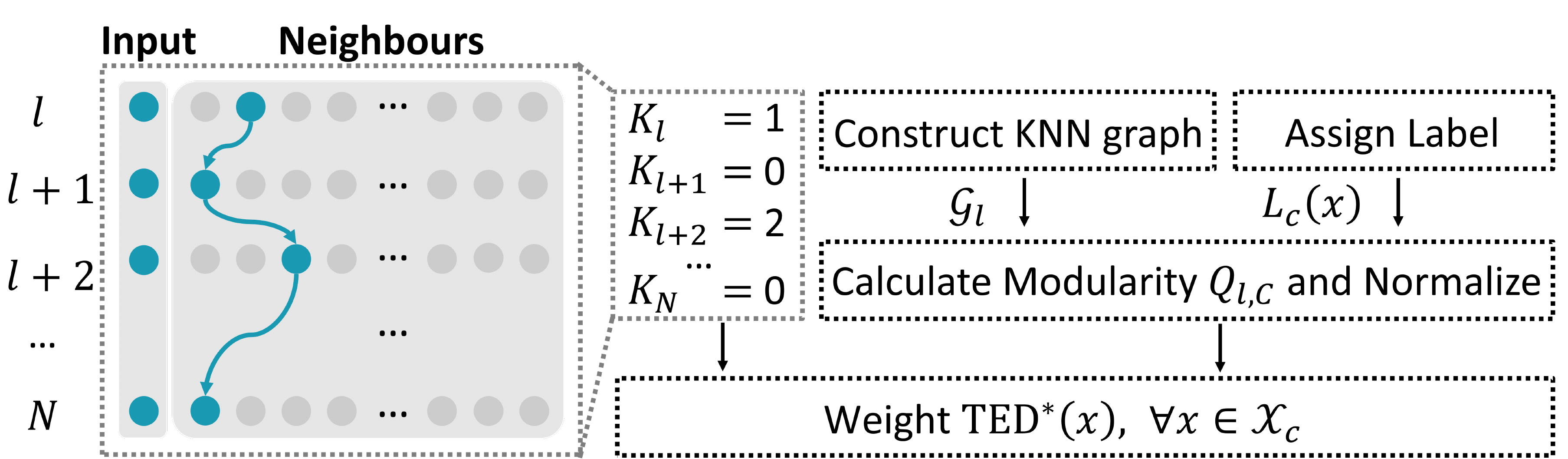} 
\caption{Overview of \ourdefense adaptive feature vector calculation structure.}
\label{fig:ted-last-structure}
\end{figure}

In this section, we provide the details of our approach
to detecting backdoor attacks. The framework of \ourdefense is illustrated in Fig.~\ref{fig:ted-last-structure}

\subsection{Defense Design}

$O_l(x)$ denotes the output of layer $l$ for input $x$, and $K_l(x)$ is the rank of $x$'s nearest neighbor from its predicted class at layer $l$.  For weight calculation, we first consider all possible classes. For each class $c \in \mathcal{Y}$ and each layer $l \in \{1, \ldots, N\}$, we calculate weights $w_{l,c}$ using one of the methods described in Section \ref{sec:weight_calculation}.

Next, given a sample $x$ and its predicted class $\hat{y}$, we compute \ourdefense's adaptive feature vector $\text{TED}^*(x)$, building upon the original TED feature vector (Eq.~\eqref{eq:tedfeaturevector}):
\begin{IEEEeqnarray}{l}
\hspace{-4\arraycolsep}\text{TED}^*(x) = [K_1(x) \cdot w_{1,\hat{y}}, \ldots, K_N(x) \cdot w_{N,\hat{y}}].\label{eq:ted_adaptive}
\end{IEEEeqnarray}
For each class $c \in \mathcal{Y}$, we train a dedicated PCA-based outlier detector.  This detector is trained using the set of TED* features computed from the subset of training data predicted to be of class $c$:  $\{\text{TED}^*(x) \mid x \in \mathcal{X}, \text{ and the predicted class of } x \text{ is } c\}$. The detector uses the sum of weighted Euclidean distances to the selected eigenvectors as a measure of anomaly \cite{zhao2019pyod}.  For a sample $x$, we compute the anomaly score $s(x)$ as the sum of weighted Euclidean distances between the sample (projected onto the PCA space) and each eigenvector of the PCA model corresponding to its predicted class $\hat{y}$.  The threshold $\tau_{c}$ is set as the $\alpha$-quantile of the anomaly scores within the set of training samples predicted as class $c$.

During the inference phase, for a new sample $x$ with predicted class $\hat{y}$, we first compute $\text{TED}^*(x)$. We then calculate the anomaly score $s(x)$ using the PCA-based outlier detector trained for class $\hat{y}$. The sample is classified as anomalous if its score exceeds the corresponding threshold $\tau_{\hat{y}}$. Algorithm~\ref{algo:modularity-ted} provides details of the modularity-based \ourdefense process.

\subsection{Weight Calculation}
\label{sec:weight_calculation}

We first construct a graph $\mathcal{G}_l$ for each layer $l$ using k-nearest neighbors (KNN) on the layer's activations $\{O_l(x) \mid x \in \mathcal{X}\}$, and the number of neighbors is set to $\sqrt{|\mathcal{X}|}$. We then assign labels $C_c(x) = 0$ if $\hat{y} = c$, and 1 otherwise. 
For each layer $l$ and class $c \in \mathcal{Y}$, we compute modularity $Q_{l,c}$ on $\mathcal{G}_l$ using $C_c(x)$:
\begin{IEEEeqnarray}{rCl}
Q_{l,c} = \sum_{i=1}^{n_c} \left[m E_i - \gamma \left(\frac{k_i}{2m}\right)^2 \right],
\label{eq:modularity}
\end{IEEEeqnarray}
where $n_c$ is the number of communities (in this case, 2), $m$ is the total number of edges in $\mathcal{G}_l$, $E_i$ is the number of edges within community $i$, $k_i$ is the sum of node degrees in community $i$, and $\gamma$ is the resolution parameter (default 1). 
We then normalize the weights across all layers for each class:
\begin{IEEEeqnarray}{rCl}
w_{l,c} = \frac{Q_{l,c} - Q_{\min,c}}{Q_{\max,c} - Q_{\min,c}},
\end{IEEEeqnarray}
where $Q_{\min,c}$ and $Q_{\max,c}$ are the minimum and maximum modularity values across all layers for the class $c$.

\begin{algorithm}[h]
\caption{Modularity-based \ourdefense 
}\label{algo:modularity-ted}

\textbf{Input:} Set of samples $\mathcal{X}$ with predicted labels $\hat{y}$ for each $x \in \mathcal{X}$, set of classes $\mathcal{Y}$, quantile $\alpha$ for threshold, total number of layers $N$, output of layer $l$ for input $x$ as $O_l(x)$

\tcc{\textcolor{color2}{Preprocessing}}
\For{each layer \(l \in \{1, \ldots, N\}\)}{
Construct graph \(\mathcal{G}_l\) using KNN ($k = \sqrt{|\mathcal{X}|}$) on \(\{O_l(x) \mid x \in \mathcal{X}\}\)

Compute \(K_l(x)\): Rank of \(x\)'s nearest neighbor from its predicted class at layer \(l\)
}
\tcc{\textcolor{color4}{Training Phase}}
\For{each class $c \in \mathcal{Y}$}{
Assign labels: $C_c(x) = \begin{cases} 0 & \text{if } \hat{y} = c \\ 1 & \text{otherwise} \end{cases}$

\For{each layer $l \in {1, \ldots, N}$}{
Compute Modularity $Q_{l,c}$ on $\mathcal{G}_l$ using $C_c(x)$ according to Eq.~\eqref{eq:modularity}
}

Normalize weights: $w_{l,c} = \frac{Q_{l,c} - Q_{\min,c}}{Q_{\max,c} - Q_{\min,c}}$ for all $l$

Compute \(\text{TED}^*(x) = [K_1(x) \cdot w_{1,c}, \ldots, K_N(x) \cdot w_{N,c}]\)

Train PCA-based outlier detector for class \(c\) on \(\{\text{TED}^*(x) \mid x \in \mathcal{X}, \hat{y} = c\}\) \\
Set threshold \(\tau_c\) as the \(\alpha\)-quantile of \(\{s(x) \mid x \in \mathcal{X}, \hat{y} = c\}\)
}

\tcc{\textcolor{color3}{Inference Phase}}
\SetKwFunction{FDetect}{Detect}
\SetKwProg{Fn}{Function}{:}{}
\Fn{\FDetect{$x$, $\hat{y}$}}{
Compute \(\text{TED}^*(x) = [K_1(x) \cdot w_{1,\hat{y}}, \ldots, K_N(x) \cdot w_{N,\hat{y}}]\) 
Compute anomaly score $s(x)$ using PCA-based outlier detector for class $\hat{y}$ \\
\Return $s(x) > \tau_{\hat{y}}$ ? ANOMALOUS : NORMAL 
}

\end{algorithm}

\subsection{Evaluation Metrics for Sample-Level Backdoor Detection}

In our evaluation, we primarily employ two common metrics as our previous work \cite{10646679}: precision and F1 score. For the defender, malicious data is considered positive, while clean data is considered negative.
Precision measures the proportion of correctly identified malicious inputs among all inputs flagged as malicious. This metric reflects the accuracy of the detection system, particularly its ability to reduce false positives.
The F1 score provides a balanced measure of model performance by combining precision and true positive rate (TPR, also known as recall).

To determine the detection thresholds, we analyze a ranking sequences of normal inputs for each class. The class-specific threshold $\tau_i$ is determined using a reject parameter $\alpha$ through PCA-based outlier detection on normal input samples. We set $\alpha = 0.05$ to achieve a 5\% false positive rate (FPR), meaning that 5\% of samples with the most deviant projections onto principal components are flagged as potential outliers. This FPR level is considered acceptable in practical applications.
Additionally, to provide a more complete performance assessment across different FPR thresholds, we include AUROC as a supplementary metric in our ablation study.

\section{Experiments}
\label{sec:Design of Backdoor Attacks and Experimental Setup}

In this section, we conduct our experiments to various adaptive attack scenarios, including the Enhanced Adaptive Attack we propose. Results for \ourdefense against two additional non-adaptive backdoor attacks are presented in Appendix~\ref{appendix:Non-Adaptive Attack}. 
Following the same setting as our previous work~\cite{10646679}, \ourdefense uses 200 clean samples per class from CIFAR-10 and GTSRB dataset, and utilizes all outputs from the Conv2D and Linear layers. For implementation, we employ the open-source Python Outlier Detection (PyOD) library \cite{zhao2019pyod}.

\subsection{Robustness Against Adaptive Attack}
\label{subsec:againstadap} 
To evaluate \ourdefense against adaptive attacks, we follow the implementation in \cite{qi2022revisiting}, training ResNet-20 models on CIFAR-10 and GTSRB datasets under both Adap-Blend and Adap-Patch scenarios. Our evaluation uses 1000 poisoned and 1000 clean samples from each dataset. As shown in Tables~\ref{tab:CIFAR-10Adaptive} and \ref{tab:gtsrbAdaptive}, \ourdefense maintains robust detection capabilities with precision $\geq$ 94.0\% and F1 score $\geq$  92.9\% across all configurations. Compared to TED, \ourdefense demonstrates superior performance, achieving 35\% higher F1 score for Adap-Blend attacks on CIFAR-10 and 63\% higher for Adap-Patch attacks on GTSRB.




\begin{table*}[h]
\centering
\renewcommand{\arraystretch}{1.2}
\caption{Comparison of detection (Precision and F1 Score, in \%) for different defenses against Adap-Blend and Adap-Patch attacks on CIFAR-10.}
\label{tab:CIFAR-10Adaptive}
\begin{tabular}{
    l
    S[table-format=2.1] S[table-format=2.1]
    S[table-format=2.1] S[table-format=2.1]
    S[table-format=2.1] S[table-format=2.1]
    S[table-format=2.1] S[table-format=2.1]
    S[table-format=2.1] S[table-format=2.1]
}
\toprule
& \multicolumn{2}{c}{STRIP} & \multicolumn{2}{c}{SCAn} & \multicolumn{2}{c}{TeCo} & \multicolumn{2}{c}{TED} & \multicolumn{2}{c}{\ourdefense} \\
\cmidrule(lr){2-3} \cmidrule(lr){4-5} \cmidrule(lr){6-7} \cmidrule(lr){8-9} \cmidrule(lr){10-11}
& {Precision} & {F1 Score} & {Precision} & {F1 Score} & {Precision} & {F1 Score} & {Precision} & {F1 Score} & {Precision} & {F1 Score} \\
\midrule
Adap-Blend & 47.4 & 8.2 & 0 & Null & 90.8 & 64.6 & 87.6 & 69.4 & \textbf{94.0} & \textbf{93.7} \\
Adap-Patch & 16.7 & 1.9 & 0 & Null & 22.1 & 2.5 & 91.4 & 79.3 & \textbf{94.8} & \textbf{92.9} \\
\bottomrule
\end{tabular}
\end{table*}

\begin{table*}[h]
\centering
\renewcommand{\arraystretch}{1.2}
\caption{Comparison of detection (Precision and F1 Score, in \%) for different defenses against Adap-Blend and Adap-Patch attacks on GTSRB.}
\label{tab:gtsrbAdaptive}
\begin{tabular}{
    l
    S[table-format=2.1] S[table-format=2.1]
    S[table-format=2.1] S[table-format=2.1]
    S[table-format=2.1] S[table-format=2.1]
    S[table-format=2.1] S[table-format=2.1]
    S[table-format=2.1] S[table-format=2.1]
}
\toprule
& \multicolumn{2}{c}{STRIP} & \multicolumn{2}{c}{SCAn} & \multicolumn{2}{c}{TeCo} & \multicolumn{2}{c}{TED} & \multicolumn{2}{c}{TED-LaST} \\
\cmidrule(lr){2-3} \cmidrule(lr){4-5} \cmidrule(lr){6-7} \cmidrule(lr){8-9} \cmidrule(lr){10-11}
& {Precision} & {F1 Score} & {Precision} & {F1 Score} & {Precision} & {F1 Score} & {Precision} & {F1 Score} & {Precision} & {F1 Score} \\
\midrule
Adap-Blend & 56.9 & 11.8 & 30.3 & 1.9 & 16.8 & 1.9 & 95.0 & 95.4 & \textbf{96.1} & \textbf{96.7} \\
Adap-Patch & 42.5 & 6.8 & 88.6 & 50.8 & 50.6 & 8.9 & 92.5 & 60.1 & \textbf{96.2} & \textbf{98.0} \\
\bottomrule
\end{tabular}
\end{table*}

\begin{figure}[t!]
\centering
    \includegraphics[width=0.25\textwidth]{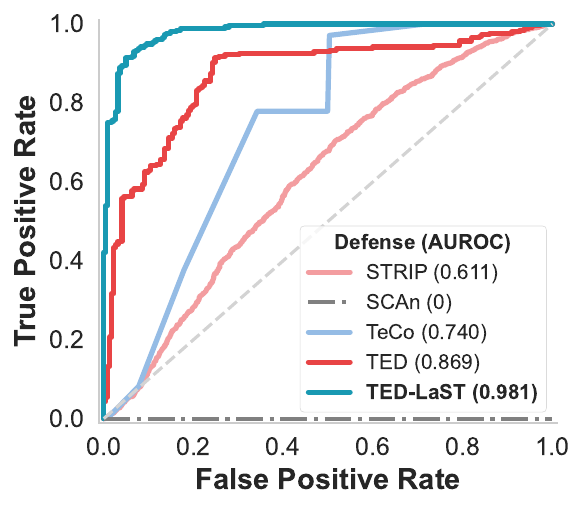}
    \caption{Defenses against Adap-Blend adaptive attack ROC curve comparison on CIFAR-10.}
    \label{fig:roc_comparison}
\end{figure}

\subsection{Evaluation on Enhanced Adaptive Attacks}

Beyond SS+L+SR and SS\&TA+L+SR as Enhanced Adaptive Attacks, we aim to understand the robustness of \ourdefense under different combinations of adaptive poisoning tricks. We examine three different trigger-to-target mapping scenarios: Basic, Source-Specific  \targetmapping (SS), and Source-Specific\&Trigger Attribute \targetmapping (SS\&TA), where Basic refers to cases where the trigger's target remains static regardless source class or trigger attributes.

\subsubsection{Settings}

For our baseline attack configuration, we implement a 6$\times$6 square trigger in the bottom right corner of input images \cite{xue2020one}, with a poison rate of 0.01.
For \laundry implementation, we follow methodologies from previous studies \cite{10646679,tang2021demon,qi2022revisiting}, selecting training samples from non-victim classes, applying triggers, and labeling them with genuine labels. The number of \laundry samples matches the number of poisoned samples.
For \slowrelease implementation, due to the square trigger being too small for further partitioning, we follow \cite{qi2022revisiting} and use a "Hello Kitty" trigger sized equal to the input. Following \cite{qi2022revisiting,xie2020dba,xue2020one}, the trigger is divided into 16 segments, with random parts integrated into training samples, where each poisoned sample randomly applies half of these segments, while the complete trigger is used for testing.
For \targetmapping implementation, we follow the setup described in \cite{xue2020one}, where two source classes and two distinct target classes are selected for the SS scenario. For the SS\&TA scenario, to balance the ASR across different classes considering varying trigger intensities, two poison rates (\textit{e.g.}, 0.1 and 0.08) are used to poison a single class, with different trigger densities (\textit{e.g.}, 0.4 and 0.6 for the square trigger, and 0.15 and 0.3 for the "Hello Kitty" trigger in \slowrelease scenarios) as per the configurations in \cite{xue2020one} and \cite{qi2022revisiting}, respectively. These varying trigger intensities ensure that poisoned samples achieve their intended target classes, respectively.

These backdoor attacks are trained on CIFAR-10 and GTSRB datasets. CIFAR-10 \cite{CIFAR10} consists of 60,000 $32 \times 32$ color images across ten classes, while GTSRB \cite{stallkamp2012man} features over 50,000 $32 \times 32$ traffic sign images across 43 classes. Both experiments employ the ResNet-20 model \cite{he2016deep}, following the setup in \cite{qi2022revisiting}.
Table~\ref{tab:AttackTechniques} illustrates the performance of our proposed Enhanced Adaptive Attacks including SS\&TA+L+SR and SS+L+SR on CIFAR-10 and GTSRB datasets, demonstrating adequate Attack Success Rate (ASR) and Clean Accuracy (Clean ACC) across the tested configurations.

\begin{table*}[ht!]
\centering
\caption{Performance of Various Adaptive Attacks with Tricks on CIFAR-10 and GTSRB Datasets. 
Settings: B (Basic), L (\laundry), SR (\slowrelease), SS (Source-Specific), TA (Trigger Attribute)}
\label{tab:AttackTechniques}
\begin{tabular}{lccccccc}
\toprule
 \multirow{2}{*}{Setting} & \multirow{2}{*}{\targetmapping} & \multirow{2}{*}{\laundry} & \multirow{2}{*}{\slowrelease} & \multicolumn{2}{c}{CIFAR-10} & \multicolumn{2}{c}{GTSRB} \\
\cmidrule(lr){5-6} \cmidrule(lr){7-8}
& & & & ASR (\%) & Clean ACC (\%) & ASR (\%) & Clean ACC (\%) \\
\midrule
B & - & - & - & 99.8  & 86.4 & 100 & 92.9 \\
L & - & $\checkmark$ & - & 100   & 80.2 & 100 & 94.9 \\
SR & - & - & $\checkmark$ & 100   & 83.0 & 100 & 96.2 \\
L+SR & - & $\checkmark$ & $\checkmark$ & 69.7  & 82.6 & 100 & 97.2 \\
\midrule
SS & SS & - & - & 97.5  & 77.9 & 100 & 95.7 \\
SS+L & SS & $\checkmark$ & - & 99.8  & 80.7 & 100 & 100 \\
SS+SR & SS & - & $\checkmark$ & 76.9  & 83.3 & 100 & 96.8 \\
SS+L+SR & SS & $\checkmark$ & $\checkmark$ & 79.0  & 79.1 & 100 & 96.9 \\
\midrule
SS\&TA & SS\&TA & - & - & 100   & 81.0 & 100 & 95.7 \\
SS\&TA+L & SS\&TA & $\checkmark$ & - & 100   & 80.3 & 100 & 100 \\
SS\&TA+SR & SS\&TA & - & $\checkmark$ & 78.3  & 83.2 & 100 & 97.7 \\
SS\&TA+L+SR & SS\&TA & $\checkmark$ & $\checkmark$ & 84.1  & 78.8 & 100 & 99.8 \\
\bottomrule
\end{tabular}
\end{table*}

\begin{table*}[htbp]
\centering
\caption{Performance (Precision and F1 Score, in \%) of different defenses against attacks Basic \targetmapping on CIFAR-10 using ResNet-20. 
Settings: B (Basic), L (\laundry), SR (\slowrelease)}
\label{tab:CIFAR-10_no_target_mapping}
\begin{tabular}{@{}lcccccccccc@{}}
\toprule
\multirow{2}{*}{Setting} & \multicolumn{2}{c}{STRIP} & \multicolumn{2}{c}{SCAn} & \multicolumn{2}{c}{TeCo} & \multicolumn{2}{c}{TED} & \multicolumn{2}{c}{TED-LaST} \\
\cmidrule(lr){2-3} \cmidrule(lr){4-5} \cmidrule(lr){6-7} \cmidrule(lr){8-9} \cmidrule(lr){10-11}
 & Precision & F1 Score & Precision & F1 Score & Precision & F1 Score & Precision & F1 Score & Precision & F1 Score \\
\midrule
B    & 95.3 & 97.3 & 94.1 & 86.5 & 74.5 & 29.1 & 94.8 & 93.1 & \textbf{96.7} & \textbf{98.0} \\
L    & 63.6 & 7.9  & 91.0 & 64.8 & 87.1 & 47.9 & \textbf{94.1} & 88.2 & 92.7 & \textbf{90.3} \\
SR   & 20.0 & 3.0  & 89.3 & 58.2 & 16.7 & 1.9  & 96.3 & 97.4 & \textbf{98.9} & \textbf{99.5} \\
L+SR & 50.9 & 10.4 & 91.0 & 64.8 & 70.3 & 20.6 & 93.0 & 92.2 & \textbf{96.1} & \textbf{97.8} \\
\bottomrule
\end{tabular}
\end{table*}

\begin{table*}[htbp]
\centering
\caption{Performance (Precision and F1 Score, in \%) of different defenses against attacks with SS (Source-Specific) \targetmapping on CIFAR-10 using ResNet-20. 
Settings: SS (Source-Specific), L (\laundry), SR (\slowrelease)}
\label{tab:CIFAR-10_sc}
\begin{tabular}{@{}lcccccccccc@{}}
\toprule
\multirow{2}{*}{Setting} & \multicolumn{2}{c}{STRIP} & \multicolumn{2}{c}{SCAn} & \multicolumn{2}{c}{TeCo} & \multicolumn{2}{c}{TED} & \multicolumn{2}{c}{TED-LaST} \\
\cmidrule(lr){2-3} \cmidrule(lr){4-5} \cmidrule(lr){6-7} \cmidrule(lr){8-9} \cmidrule(lr){10-11}
 & Precision & F1 Score & Precision & F1 Score & Precision & F1 Score & Precision & F1 Score & Precision & F1 Score \\
\midrule
SS      & 46.2 & 8.7  & 85.5 & 44.0 & 20.0 & 1.9 & \textbf{95.4} & 93.9 & 94.1 & \textbf{95.5} \\
SS+L    & \textbf{94.4} & \textbf{94.6} & 83.3 & 38.5 & 18.1 & 1.9 & 90.8 & 83.3 & 93.3 & \textbf{94.6} \\
SS+SR   & 28.6 & 4.6  & 83.2 & 38.1 & 18.1 & 1.8 & \textbf{94.7} & 75.3 & 92.9 & \textbf{88.0} \\
SS+L+SR & 58.7 & 13.1 & 82.9 & 37.5 & 16.7 & 1.9 & 88.0 & 67.1 & \textbf{93.1} & \textbf{91.0} \\
\bottomrule
\end{tabular}
\end{table*}

\begin{table*}[htbp]
\centering
\caption{Performance (Precision and F1 Score, in \%) of different defenses against attacks with SS\&TA (Source-Specific\&Trigger Attribute) \targetmapping on CIFAR-10 under ResNet-20. 
Settings: SS\&TA (Source-Specific\&Trigger Attribute), L (\laundry), SR (\slowrelease)}
\label{tab:CIFAR-10_sc_ta}
\begin{tabular}{@{}lcccccccccc@{}}
\toprule
\multirow{2}{*}{Setting} & \multicolumn{2}{c}{STRIP} & \multicolumn{2}{c}{SCAn} & \multicolumn{2}{c}{TeCo} & \multicolumn{2}{c}{TED} & \multicolumn{2}{c}{TED-LaST} \\
\cmidrule(lr){2-3} \cmidrule(lr){4-5} \cmidrule(lr){6-7} \cmidrule(lr){8-9} \cmidrule(lr){10-11}
 & Precision & F1 Score & Precision & F1 Score & Precision & F1 Score & Precision & F1 Score & Precision & F1 Score \\
\midrule
SS\&TA    & 83.3 & 17.9 & 85.3 & 42.7 & 10.2 & 1.7  & \textbf{96.5} & 90.7 & 94.0 & \textbf{92.8} \\
SS\&TA+L  & 84.4 & 44.7 & 81.3 & 34.4 & 25.0 & 4.9  & 91.2 & 86.5 & \textbf{93.4} & \textbf{89.9} \\
SS\&TA+SR & 12.1 & 1.5  & 81.9 & 35.4 & 16.4 & 1.9  & 89.4 & 58.7 & \textbf{92.0} & \textbf{85.6} \\
SS\&TA+L+SR & 46.6 & 11.9 & 85.2 & 43.0 & 16.7 & 1.9 & 89.5 & 76.4 & \textbf{91.5} & \textbf{87.9} \\
\bottomrule
\end{tabular}
\end{table*}

\begin{table*}[htbp]
\centering
\caption{Performance (Precision and F1 Score, in \%) of different defenses against Basic \targetmapping attacks on GTSRB under ResNet-20. 
Settings: B (Basic), L (\laundry), SR (\slowrelease)}
\label{tab:gtsrb_no_target_mapping}
\begin{tabular}{@{}lcccccccccc@{}}
\toprule
\multirow{2}{*}{Setting} & \multicolumn{2}{c}{STRIP} & \multicolumn{2}{c}{SCAn} & \multicolumn{2}{c}{TeCo} & \multicolumn{2}{c}{TED} & \multicolumn{2}{c}{TED-LaST} \\
\cmidrule(lr){2-3} \cmidrule(lr){4-5} \cmidrule(lr){6-7} \cmidrule(lr){8-9} \cmidrule(lr){10-11}
 & Precision & F1 Score & Precision & F1 Score & Precision & F1 Score & Precision & F1 Score & Precision & F1 Score \\
\midrule
B    & 91.6 & 87.3 & 90.9 & 79.3 & 89.9 & 57.7 & \textbf{95.1} & \textbf{97.0} & \textbf{95.1} & \textbf{97.0} \\
L    & 79.7 & 21.8 & 89.2 & 56.7 & 63.2 & 15.1 & \textbf{95.5} & \textbf{97.7} & \textbf{95.5} & \textbf{97.7} \\
SR   & 3.3  & 0.4  & 91.0 & 64.9 & 88.0 & 51.8 & 94.6 & 97.2 & \textbf{95.5} & \textbf{97.7} \\
L+SR & 5.3  & 0.4  & 89.8 & 59.1 & 78.0 & 29.4 & 94.6 & 97.2 & \textbf{95.5} & \textbf{97.7} \\
\bottomrule
\end{tabular}
\end{table*}

\begin{table*}[htbp]
\centering
\caption{Performance (Precision and F1 Score, in \%) of different defenses against attacks with SS (Source-Specific) \targetmapping on GTSRB under ResNet-20. 
Settings: SS (Source-Specific), L (\laundry), SR (\slowrelease)}
\label{tab:gtsrb_source_class}
\begin{tabular}{@{}lcccccccccc@{}}
\toprule
\multirow{2}{*}{Setting} & \multicolumn{2}{c}{STRIP} & \multicolumn{2}{c}{SCAn} & \multicolumn{2}{c}{TeCo} & \multicolumn{2}{c}{TED} & \multicolumn{2}{c}{TED-LaST} \\
\cmidrule(lr){2-3} \cmidrule(lr){4-5} \cmidrule(lr){6-7} \cmidrule(lr){8-9} \cmidrule(lr){10-11}
 & Precision & F1 Score & Precision & F1 Score & Precision & F1 Score & Precision & F1 Score & Precision & F1 Score \\
\midrule
SS    & 45.7 & 11.2 & 84.6 & 41.5 & 65.0 & 13.6 & 94.7 & 97.3 & \textbf{94.9} & \textbf{97.4} \\
SS+L  & 15.6 & 1.9  & 94.8 & 93.1 & 78.0 & 29.4 & 94.5 & 97.2 & \textbf{94.9} & \textbf{97.4} \\
SS+SR & 3.3  & 0.4  & 79.1 & 30.5 & 87.4 & 49.7 & 95.9 & 95.2 & \textbf{96.0} & \textbf{96.6} \\
SS+L+SR & 3.2  & 0.4  & 95.0 & 95.2 & 56.3 & 12.0 & 94.6 & 95.1 & \textbf{94.8} & \textbf{97.2} \\
\bottomrule
\end{tabular}
\end{table*}

\begin{table*}[htbp]
\centering
\caption{Performance (Precision and F1 Score, in \%) of different defenses against attacks with SS\&TA (Source-Specific \& Trigger-Attribute) \targetmapping on GTSRB under ResNet-20. 
Settings: SS\&TA (Source-Specific\&Trigger Attribute), L (\laundry), SR (\slowrelease)}
\label{tab:gtsrb_sc_ta}
\begin{tabular}{@{}lcccccccccc@{}}
\toprule
\multirow{2}{*}{Setting} & \multicolumn{2}{c}{STRIP} & \multicolumn{2}{c}{SCAn} & \multicolumn{2}{c}{TeCo} & \multicolumn{2}{c}{TED} & \multicolumn{2}{c}{TED-LaST} \\
\cmidrule(lr){2-3} \cmidrule(lr){4-5} \cmidrule(lr){6-7} \cmidrule(lr){8-9} \cmidrule(lr){10-11}
 & Precision & F1 Score & Precision & F1 Score & Precision & F1 Score & Precision & F1 Score & Precision & F1 Score \\
\midrule
SS\&TA    & 4.5 & 0.4 & 74.5 & 24.4 & 87.4 & 49.7 & \textbf{94.3} & \textbf{97.1} & \textbf{94.3} & \textbf{97.1} \\
SS\&TA+L  & 67.1 & 16.5 & 95.0 & 95.3 & 16.9 & 1.9 & 94.9 & 96.7 & \textbf{95.9} & \textbf{97.9} \\
SS\&TA+SR & 31.0 & 6.5 & 84.3 & 40.7 & 16.7 & 1.9 & 95.3 & 97.6 & \textbf{95.7} & \textbf{97.8} \\
SS\&TA+L+SR & 2.9 & 0.4 & 94.9 & 94.7 & 17.5 & 1.9 & 95.4 & 95.6 & \textbf{95.7} & \textbf{97.7} \\
\bottomrule
\end{tabular}
\end{table*}

\begin{table}[]
\centering
\caption{Attack Performance and TED-LaST Effectiveness on ImageNet100}
\label{table:AttackDefensePerformance}
\setlength{\tabcolsep}{4pt}
\begin{tabular}{lcccccc}
\toprule
& \multicolumn{2}{c}{Attack Performance} & \multicolumn{2}{c}{TED-LaST Performance} \\
\cmidrule(lr){2-3} \cmidrule(lr){4-5}
Setting & ASR (\%) & Clean ACC (\%) & Precision (\%) & F1 Score (\%) \\
\midrule
B & 100 & 83.2 & 93.0 & 92.5 \\
L & 100 & 82.4 & 95.4 & 93.1 \\
SR & 100 & 83.4 & 95.0 & 95.0 \\
L+SR & 87.0 & 78.5 & 94.2 & 87.1 \\
\midrule
SS & 99.8 & 82.4 & 95.8 & 95.5 \\
SS+L & 100 & 82.7 & 94.4 & 95.4 \\
SS+SR & 81.3 & 84.2 & 95.0 & 97.2 \\
SS+L+SR & 88.8 & 82.3 & 94.8 & 95.5 \\
\midrule
SS\&TA & 100 & 80.1 & 96.2 & 98.1 \\
SS\&TA+L & 100 & 81.1 & 94.7 & 95.1 \\
SS\&TA+SR & 94.8 & 84.1 & 91.9 & 89.9 \\
SS\&TA+L+SR & 88.8 & 83.6 & 94.8 & 97.2 \\
\bottomrule
\end{tabular}
\end{table}

\subsubsection{Results}
\label{subsection:comparison-of-ourdefense-against-attacks}

Tables~\ref{tab:CIFAR-10_no_target_mapping}, \ref{tab:CIFAR-10_sc}, and \ref{tab:CIFAR-10_sc_ta} demonstrate \ourdefense's robustness against various backdoor attacks on CIFAR-10. \ourdefense consistently outperforms existing defenses including STRIP, SCAn, TeCo, and TED, especially against Enhanced Adaptive Attacks, validating our analysis in Section~\ref{Existing Defenses Against Adaptive Attacks}.

For SCAn, its performance significantly degrades against Adap-Blend attacks where attackers deliberately minimize the feature representation differences between clean and poisoned samples, challenging its distribution-based detection mechanism. 
STRIP shows inherent limitations against adaptive attacks, particularly Adap-Patch, where attackers successfully evade entropy-based detection by weakening the correlation between backdoor triggers and target labels. 
Similarly, TeCo's effectiveness diminishes when attackers deliberately blend malicious samples with clean ones in latent space, consistent with findings in~\cite{liu2023detecting}.

The effectiveness of \ourdefense becomes more pronounced with increasing attack complexity, primarily due to: (1) Latent space inseparability: Adaptive attacks induce feature space inseparability, challenging defenses relying on clear benign-malicious separations; (2) \targetmapping complexity: Sophisticated mapping patterns reduce trigger reliability as malicious indicators, particularly affecting TED's topological analysis. \ourdefense's supervised-label dynamic tracking and adaptive weighting mechanisms effectively counter these challenges. This robust performance stems from \ourdefense{}'s ability to capture subtle class-specific anomalies through supervised-label dynamic tracking, particularly effective against attacks that blur benign-malicious distinctions in global topological space.

Results on GTSRB (Tables~\ref{tab:gtsrb_no_target_mapping}, \ref{tab:gtsrb_source_class}, and \ref{tab:gtsrb_sc_ta}) further demonstrate \ourdefense's superiority. While defense efficacy generally declines as \targetmapping complexity increases from Basic to SS\&TA, \ourdefense shows minimal performance degradation. For instance, in SS+SR scenarios on GTSRB, \ourdefense maintains a 96.6\% F1 Score compared to TED's 95.2\%. Even under the most complex SS\&TA+L+SR attack on CIFAR-10, \ourdefense achieves an 87.9\% F1 Score, substantially outperforming TED's 76.4\%.


\subsection{Effectiveness of \ourdefense Against Enhanced Adaptive Attacks on Large-Scale Dataset}
\label{appendix:large scale dataset testing}
To validate the scalability and effectiveness of \ourdefense, we evaluate it on ImageNet100, a subset of ImageNet \cite{deng2009imagenet} comprising 100 classes. Our experiments use 224x224 pixel images, a ResNet101 model \cite{he2016deep}, and a resized "Hello Kitty" trigger covering the entire image.
The increased class diversity of ImageNet100 introduces higher complexity in the topological space. Benign samples navigate through a more intricate manifold, traversing larger topological distances compared to smaller datasets. This poses challenges to our defense system, requiring it to distinguish between larger natural variations and subtle shifts induced by attacks.
Despite these challenges, as shown in Table \ref{table:AttackDefensePerformance}, \ourdefense demonstrates robust performance across various Enhanced Adaptive Attack configurations. It consistently maintains high Precision rates (often exceeding 94\%) and strong F1 Scores (mostly above 90\%) across different attack settings. 
Even in the most sophisticated attack scenarios (SS\&TA+L+SR), \ourdefense achieves an F1 Score of 97.2\%, showcasing its ability to adapt to complex attack patterns.
This performance highlights \ourdefense's strong scalability for larger-scale, more complex image tasks and real-world applications.


\section{Ablation Study}
\label{sec:ablation study and instight}

\subsection{Label Information in Outlier Detection}
\label{Subsec:OutlierDetection}

\begin{figure}[t!]
\centering
\begin{tabular}{@{}c@{\hspace{1mm}}c@{\hspace{1mm}}c@{\hspace{1mm}}c@{\hspace{1mm}}c@{\hspace{1mm}}c@{}}
\includegraphics[width=0.49\linewidth]{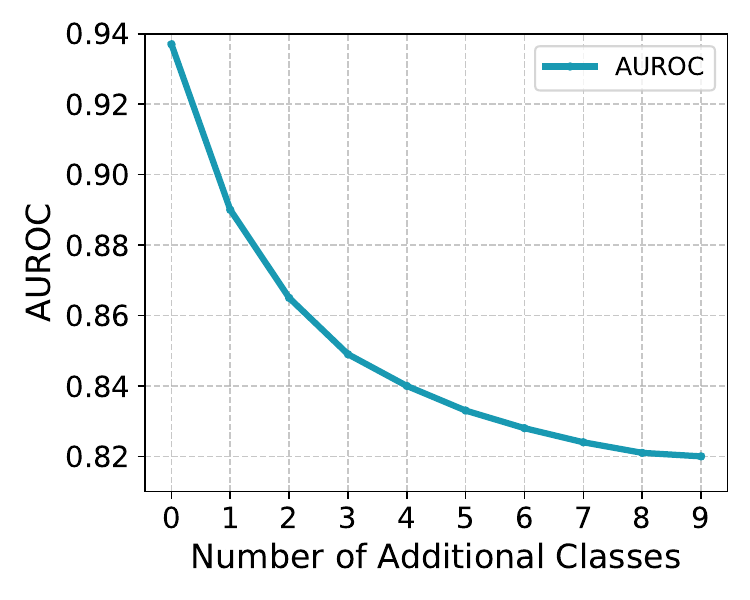} 
&
\includegraphics[width=0.49\linewidth]{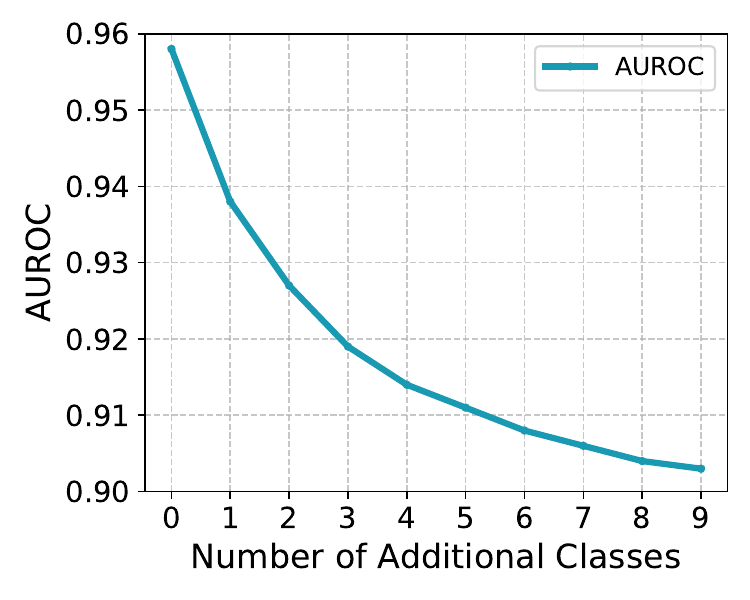} \\
(a) Adap-Blend & (b) Adap-Patch
\end{tabular}
\caption{AUROC for outlier detection with varying numbers of additional classes under (a) Adap-Blend and (b) Adap-Patch attacks on CIFAR-10. The x-axis represents the number of additional random classes included in the training data, while the y-axis shows the corresponding AUROC scores.}
\label{fig:auc_plots}
\end{figure}

This ablation study examines the importance of class-specific information in enhancing defense robustness against adaptive attacks. Fig.~\ref{fig:auc_plots} shows the effect of varying training data composition on outlier detection performance for Adap-Blend (Fig.~\ref{fig:auc_plots}a) and Adap-Patch (Fig.~\ref{fig:auc_plots}b) attacks.

Training on clean samples from the predicted class alone yields the highest AUROC scores. Including additional random classes degrades performance, with the first unrelated class causing the most significant drop. Further additions result in continued AUROC score declines at a decreasing rate.

This degradation comes from the dilution of class-specific topological features in the training data. As unrelated classes are introduced, the detector's ability to identify attack-induced deviations from expected class patterns diminishes. This observation aligns with the analysis of inseparability in topological space (Section \ref{subsection:TEDLimitationAgainstAdatptiveAttack}), where adaptive attacks blur distinctions between clean and poisoned samples.

\subsection{Adaptive Weighting Under Different CTD Thresholds}
\label{subsec:Adaptive Weighting Under Different CTD Thresholds}

We investigate the effectiveness of our modularity-based weighted method compared to the non-weighted approach across various Cumulative Topological Distance (CTD) thresholds. The CTD metric (Eq.~\eqref{eq:ctd}) quantifies the topological distance change as a sample traverses network layers. As shown in Fig.~\ref{fig:cdtplot}a and Fig.~\ref{fig:cdtplot}b, both Adap-Blend and Adap-Patch attacks generate malicious samples with varying CTD values. Samples with lower CTD values, representing more subtle perturbations, are generally more challenging to detect.

\begin{figure}[t]
\centering
\includegraphics[width=0.9\linewidth]{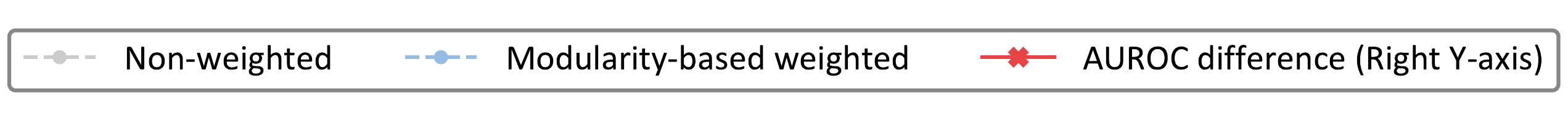}
\vspace{2mm} 
\begin{tabular}{@{}c@{\hspace{5mm}}c@{}}
\includegraphics[width=0.48\linewidth]{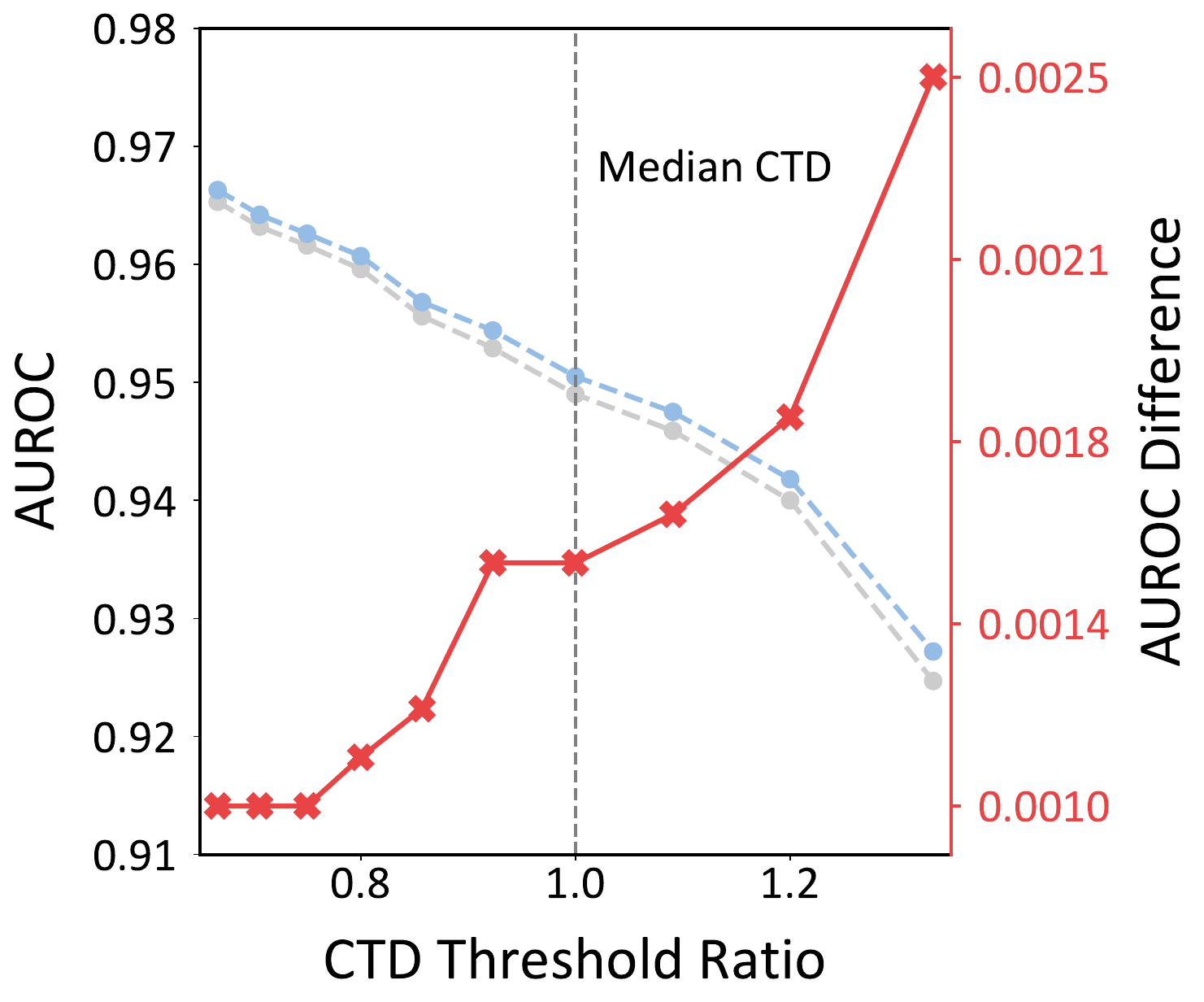} &
\includegraphics[width=0.48\linewidth]{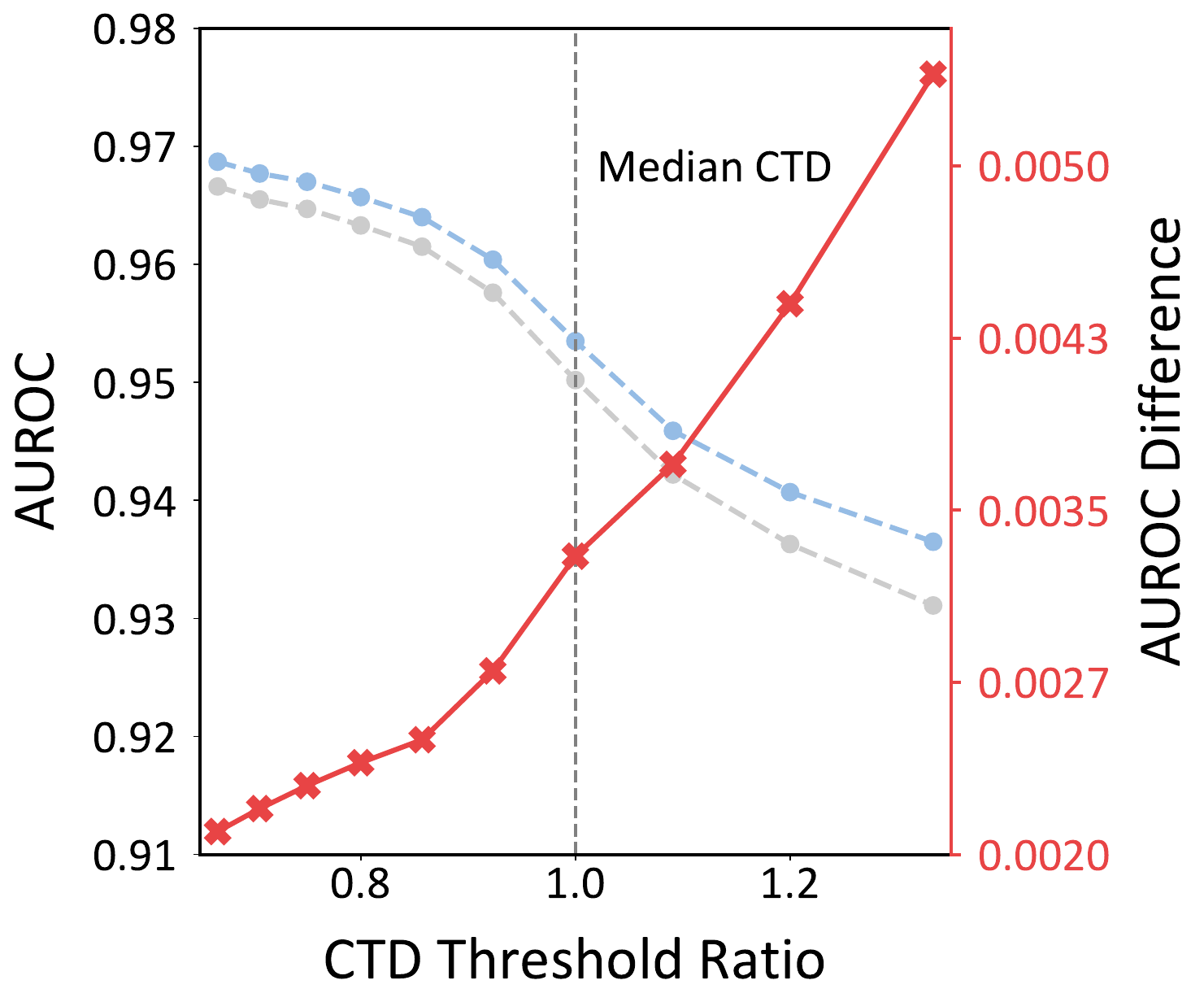} \\
(a) Adap-Blend & (b) Adap-Patch
\end{tabular}
\caption{Comparison of AUROC differences across various CTD Threshold Ratios. The modularity-based weighted method consistently shows higher AUROC than the non-weighted approach under all ratios. The gap widens as the ratio increases, indicating that the modularity-based weighted method provides more sensitivity to malicious samples with subtle perturbations.}
\label{fig:auc_CTD_plots}
\end{figure}

We define the CTD Threshold Ratio as the median CTD divided by the actual CTD value. Higher ratios correspond to lower CTD values, indicating more subtle perturbations. Our experiments use datasets with equal numbers of malicious and clean samples, filtered based on different CTD thresholds.
Fig.~\ref{fig:auc_CTD_plots} illustrates the performance difference between the modularity-weighted and non-weighted methods across CTD Threshold Ratios. 

The results reveal a clear trend: as the CTD Threshold Ratio increases (i.e., when we focus on malicious samples with lower CTD values), the modularity-based weighted method consistently outperforms the non-weighted method. This performance gap becomes more pronounced at higher CTD Threshold Ratios.

\begin{figure*}[h!]
\centering
\begin{tabular}{@{}c@{\hspace{1mm}}c@{}}
\includegraphics[width=0.48\textwidth, trim={0 0 0 130pt}, clip]{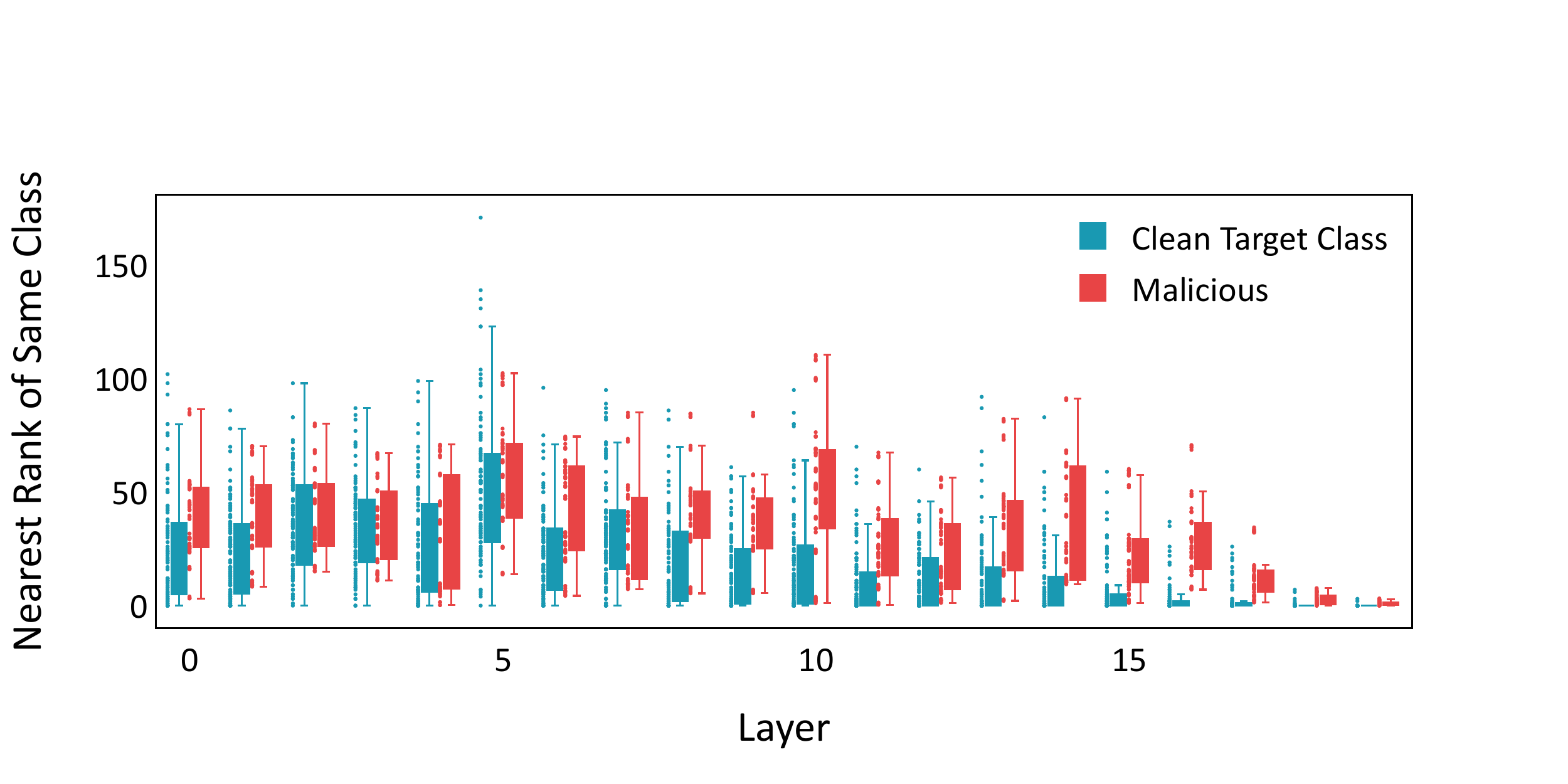} &
\includegraphics[width=0.48\textwidth, trim={0 0 0 130pt}, clip]{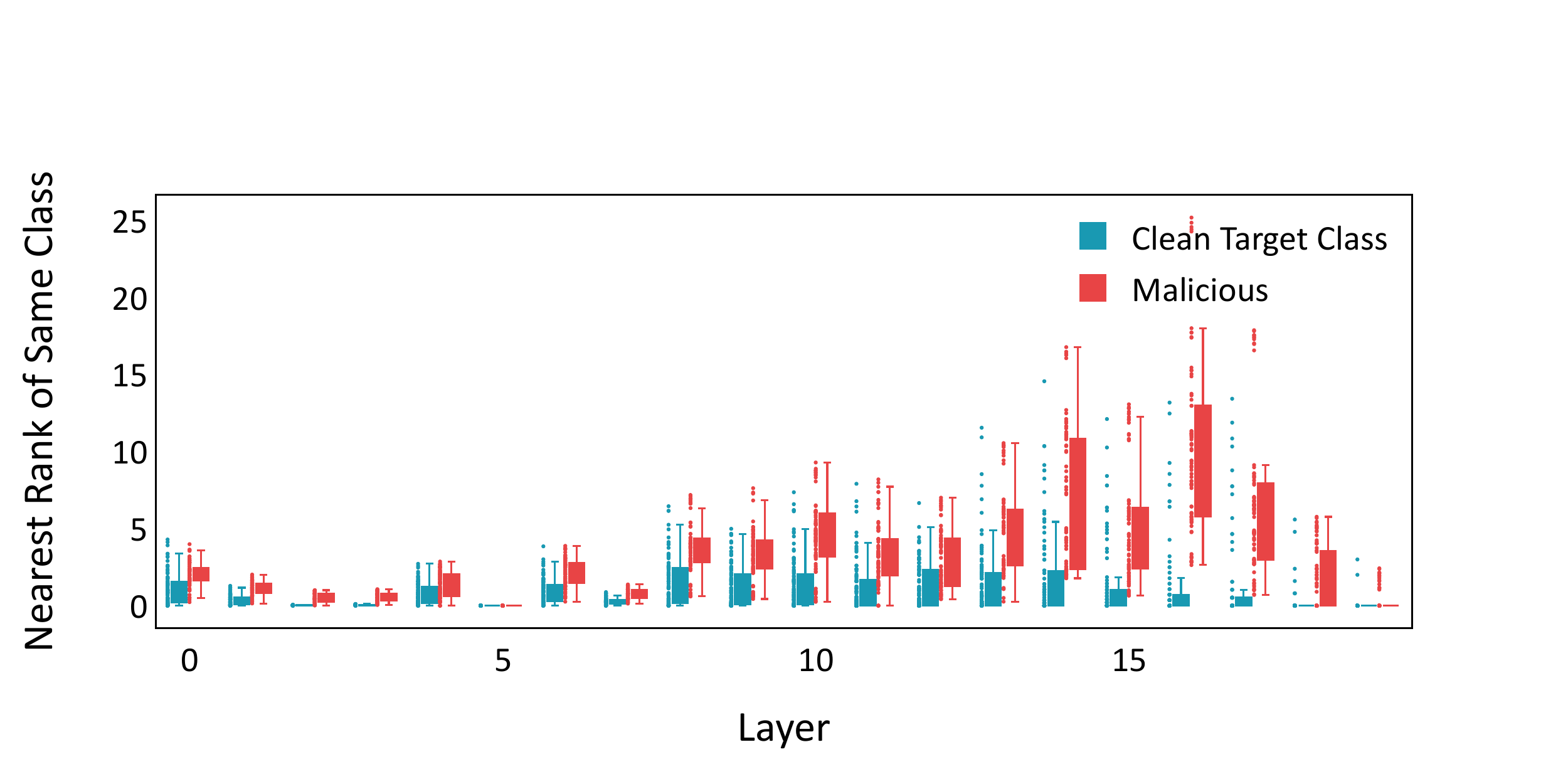} \\[-1ex]
(a) Non-weighted (target class A) & (b) Modularity-based weighted (target class A) \\
\includegraphics[width=0.48\textwidth, trim={0 0 0 130pt}, clip]{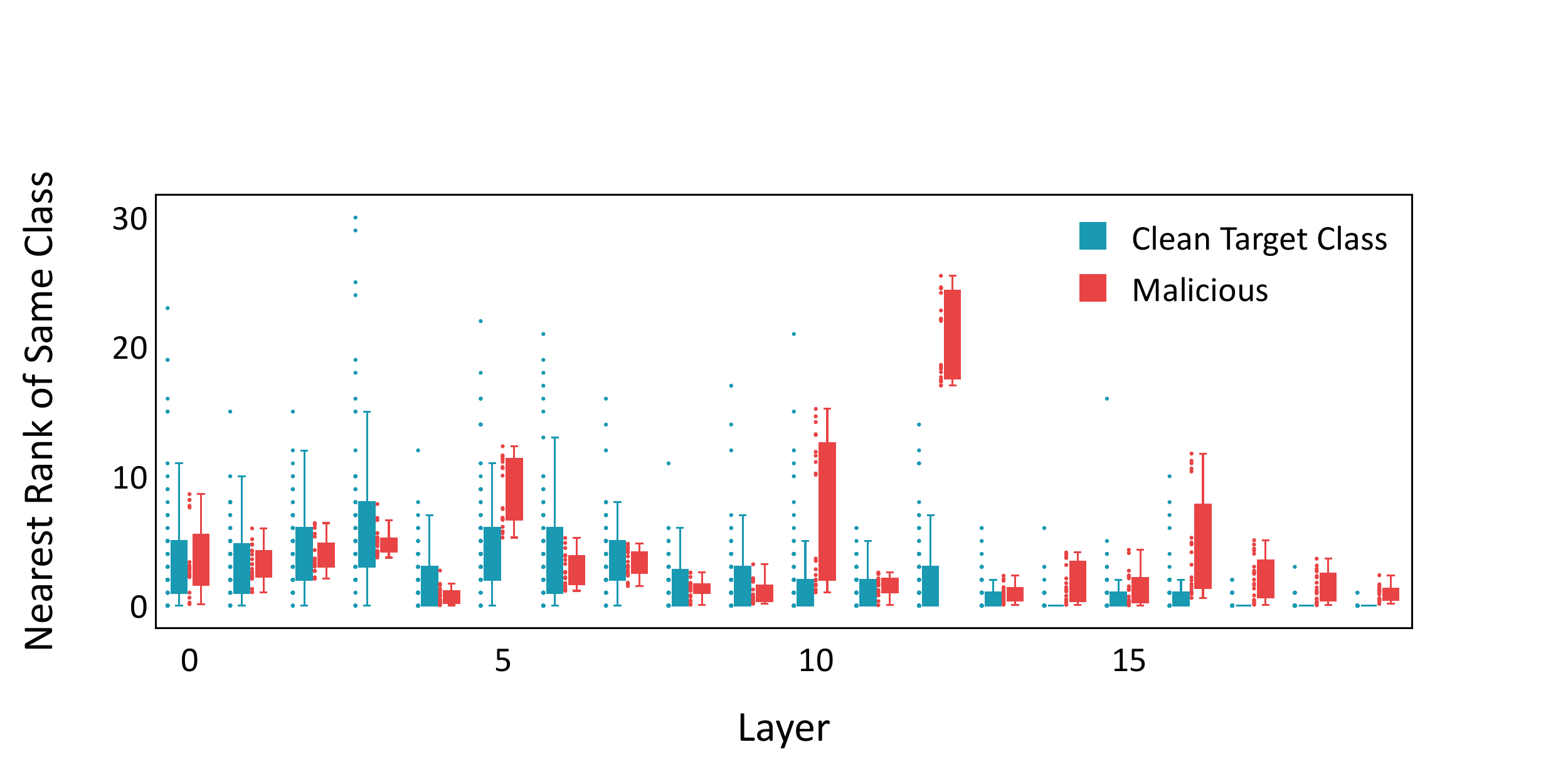} &
\includegraphics[width=0.48\textwidth, trim={0 0 0 130pt}, clip]{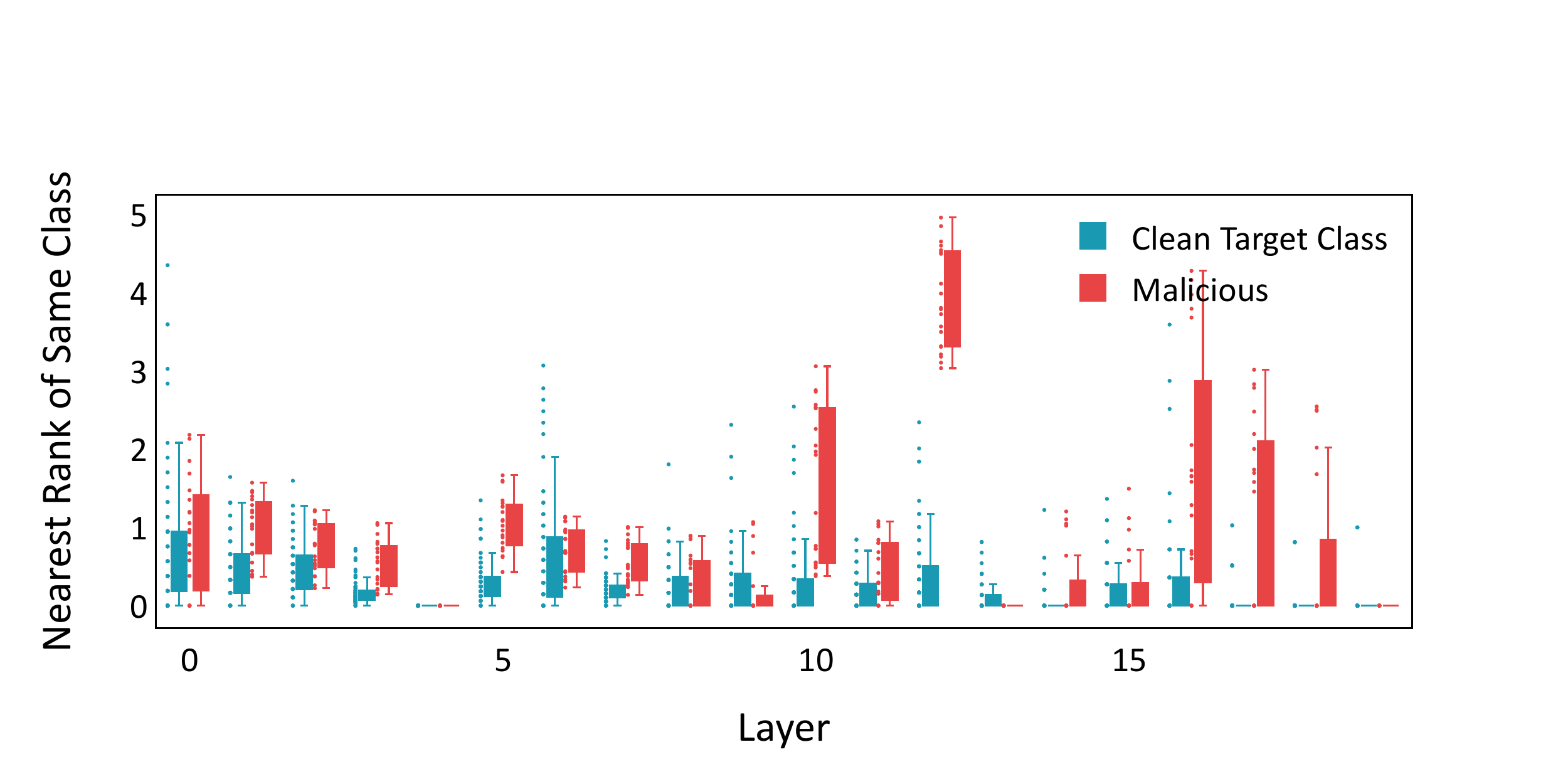} \\[-1ex]
(c) Non-weighted (target class B) & (d) Modularity-based weighted (target class B)
\end{tabular}
\caption{Box plots of topological feature vectors against SS\&TA+L+SR with dynamic targets. (a) and (c) show unweighted features for two different target classes, revealing subtle perturbations that limit discrimination. (b) and (d) demonstrate improved separation between clean and malicious samples after applying modularity-based adaptive layer emphasis for the respective target classes.}
\label{fig:ssta+l+srweightingcompare}
\end{figure*}

\subsection{Intuition Behind \ourdefense Against Adaptive Attacks}
\label{subsection:intuition}

Adaptive backdoor attacks pose a significant challenge by concealing poisoned data within the complex topological space of neural networks. As visualized in Fig.~\ref{fig:tsne_plots}a and Fig.~\ref{fig:tsne_plots}b, these attacks blur the boundaries between clean and poisoned samples.

To address these challenges, our defense strategy shifts from global analysis to a more granular approach. This strategy is based on the premise that even subtle manipulations leave detectable class-wise traces. Specifically, adaptive attacks inevitably cause deviations from benign behavior within the target class (discussed in Section~\ref{subsection:Label-Supervised Dynamics Tracking}). These deviations, while subtle, provide crucial indicators for our detection method. We also find that additional class augmentation does not enhance detection capabilities (discussed in Section~\ref{Subsec:OutlierDetection}).

Our modularity-based weighted method consistently outperforms non-weighted approaches, especially for malicious samples with subtle perturbations (Section~\ref{subsec:Adaptive Weighting Under Different CTD Thresholds}). This improved sensitivity stems from the weighting scheme's emphasis on the most informative topological features, enabling refined differentiation between benign and malicious samples.

Importantly, our adaptive weighting approach remains effective even against Enhanced Adaptive Attacks. These advanced attacks use dynamic target classes and shared triggers to deeply mix malicious samples with benign representations in the topological space. Despite this sophisticated mixing, our method can still detect the subtle perturbations introduced by these attacks (illustrated in Appendix~\ref{appendix:Subtle Perturbations on Noise Quantification and Adaptive Weighting}).

\section{Conclusion}
\label{sec:Conclusion}

This study introduces \ourdefense, a novel defense strategy against adaptive backdoor attacks in DNNs. \ourdefense leverages persistent topological perturbations on target classes and uses supervised label information to enhance differentiation between poisoned and clean samples. We implement adaptive layer emphasis to address subtle perturbations caused by these attacks.
Our approach outperforms several state-of-the-art defenses in countering SOTA adaptive attacks and our proposed Enhanced Adaptive Attacks. Future work could include further improving the defense effectiveness while exploring the application of \ourdefense in federated learning or multi-modal learning scenarios. These extensions would contribute to enhancing the security of robust machine learning.

{
\small
\bibliographystyle{IEEEtran}
\bibliography{reference}

\begin{thebibliography}{10}
\providecommand{\url}[1]{#1}
\csname url@samestyle\endcsname
\providecommand{\newblock}{\relax}
\providecommand{\bibinfo}[2]{#2}
\providecommand{\BIBentrySTDinterwordspacing}{\spaceskip=0pt\relax}
\providecommand{\BIBentryALTinterwordstretchfactor}{4}
\providecommand{\BIBentryALTinterwordspacing}{\spaceskip=\fontdimen2\font plus
\BIBentryALTinterwordstretchfactor\fontdimen3\font minus \fontdimen4\font\relax}
\providecommand{\BIBforeignlanguage}[2]{{%
\expandafter\ifx\csname l@#1\endcsname\relax
\typeout{** WARNING: IEEEtran.bst: No hyphenation pattern has been}%
\typeout{** loaded for the language `#1'. Using the pattern for}%
\typeout{** the default language instead.}%
\else
\language=\csname l@#1\endcsname
\fi
#2}}
\providecommand{\BIBdecl}{\relax}
\BIBdecl

\bibitem{9127813}
Y.~Guo, H.~Wang, Q.~Hu, H.~Liu, L.~Liu, and M.~Bennamoun, ``Deep learning for 3d point clouds: A survey,'' \emph{IEEE Transactions on Pattern Analysis and Machine Intelligence}, vol.~43, no.~12, pp. 4338--4364, 2021.

\bibitem{8585066}
T.~Afouras, J.~S. Chung, A.~Senior, O.~Vinyals, and A.~Zisserman, ``Deep audio-visual speech recognition,'' \emph{IEEE Transactions on Pattern Analysis and Machine Intelligence}, vol.~44, no.~12, pp. 8717--8727, 2022.

\bibitem{9420291}
J.~Cao, Y.~Pang, J.~Xie, F.~S. Khan, and L.~Shao, ``From handcrafted to deep features for pedestrian detection: A survey,'' \emph{IEEE Transactions on Pattern Analysis and Machine Intelligence}, vol.~44, no.~9, pp. 4913--4934, 2022.

\bibitem{gu2017badnets}
T.~Gu, B.~Dolan-Gavitt, and S.~Garg, ``Badnets: {I}dentifying vulnerabilities in the machine learning model supply chain,'' \emph{arXiv preprint arXiv:1708.06733}, 2017.

\bibitem{li2021backdoor}
L.~Li, D.~Song, X.~Li, J.~Zeng, R.~Ma, and X.~Qiu, ``Backdoor attacks on pre-trained models by layerwise weight poisoning,'' \emph{arXiv preprint arXiv:2108.13888}, 2021.

\bibitem{liu2017trojaning}
Y.~Liu, S.~Ma, Y.~Aafer, W.-C. Lee, J.~Zhai, W.~Wang, and X.~Zhang, ``Trojaning attack on neural networks,'' in \emph{25th Annual Network And Distributed System Security Symposium (NDSS 2018)}, 2018.

\bibitem{yang2021careful}
W.~Yang, L.~Li, Z.~Zhang, X.~Ren, X.~Sun, and B.~He, ``Be careful about poisoned word embeddings: Exploring the vulnerability of the embedding layers in {NLP} models,'' in \emph{ACL}, 2021, pp. 2048--2058.

\bibitem{adi2018turning}
Y.~Adi, C.~Baum, M.~Cisse, B.~Pinkas, and J.~Keshet, ``Turning your weakness into a strength: {W}atermarking deep neural networks by backdooring,'' in \emph{USENIX Security}, 2018, pp. 1615--1631.

\bibitem{lin2020composite}
J.~Lin, L.~Xu, Y.~Liu, and X.~Zhang, ``Composite backdoor attack for deep neural network by mixing existing benign features,'' in \emph{Proceedings of the 2020 ACM SIGSAC Conference on Computer and Communications Security}, 2020, pp. 113--131.

\bibitem{tang2021demon}
D.~Tang, X.~Wang, H.~Tang, and K.~Zhang, ``Demon in the variant: {S}tatistical analysis of dnns for robust backdoor contamination detection.'' in \emph{USENIX Security Symposium}, 2021, pp. 1541--1558.

\bibitem{chen2017targeted}
X.~Chen, C.~Liu, B.~Li, K.~Lu, and D.~Song, ``Targeted backdoor attacks on deep learning systems using data poisoning,'' \emph{arXiv preprint arXiv:1712.05526}, 2017.

\bibitem{zeng2021rethinking}
Y.~Zeng, W.~Park, Z.~M. Mao, and R.~Jia, ``Rethinking the backdoor attacks' triggers: A frequency perspective,'' in \emph{Proceedings of the IEEE/CVF International Conference on Computer Vision}, 2021, pp. 16\,473--16\,481.

\bibitem{quiring2020backdooring}
E.~Quiring and K.~Rieck, ``Backdooring and poisoning neural networks with image-scaling attacks,'' in \emph{2020 IEEE Security and Privacy Workshops (SPW)}.\hskip 1em plus 0.5em minus 0.4em\relax IEEE, 2020, pp. 41--47.

\bibitem{nguyen2020input}
T.~A. Nguyen and A.~Tran, ``Input-aware dynamic backdoor attack,'' \emph{Advances in Neural Information Processing Systems}, vol.~33, pp. 3454--3464, 2020.

\bibitem{li2021invisible}
Y.~Li, Y.~Li, B.~Wu, L.~Li, R.~He, and S.~Lyu, ``Invisible backdoor attack with sample-specific triggers,'' in \emph{Proceedings of the IEEE/CVF International Conference on Computer Vision}, 2021, pp. 16\,463--16\,472.

\bibitem{duan2024conditional}
Q.~Duan, Z.~Hua, Q.~Liao, Y.~Zhang, and L.~Y. Zhang, ``Conditional backdoor attack via jpeg compression,'' in \emph{Proceedings of the AAAI Conference on Artificial Intelligence}, vol.~38, no.~18, 2024, pp. 19\,823--19\,831.

\bibitem{10191661}
X.~Mo, L.~Y. Zhang, N.~Sun, W.~Luo, and S.~Gao, ``Backdoor attack on deep neural networks in perception domain,'' in \emph{2023 International Joint Conference on Neural Networks (IJCNN)}, 2023, pp. 01--08.

\bibitem{10122715}
C.~Wu, D.~Lian, Y.~Ge, Z.~Zhu, and E.~Chen, ``Influence-driven data poisoning for robust recommender systems,'' \emph{IEEE Transactions on Pattern Analysis and Machine Intelligence}, vol.~45, no.~10, pp. 11\,915--11\,931, 2023.

\bibitem{9448491}
K.~Ma, Q.~Xu, J.~Zeng, X.~Cao, and Q.~Huang, ``Poisoning attack against estimating from pairwise comparisons,'' \emph{IEEE Transactions on Pattern Analysis and Machine Intelligence}, vol.~44, no.~10, pp. 6393--6408, 2022.

\bibitem{xu2021detecting}
X.~Xu, Q.~Wang, H.~Li, N.~Borisov, C.~A. Gunter, and B.~Li, ``Detecting ai trojans using meta neural analysis,'' in \emph{2021 IEEE Symposium on Security and Privacy (SP)}.\hskip 1em plus 0.5em minus 0.4em\relax IEEE, 2021, pp. 103--120.

\bibitem{liu2019abs}
Y.~Liu, W.-C. Lee, G.~Tao, S.~Ma, Y.~Aafer, and X.~Zhang, ``Abs: Scanning neural networks for back-doors by artificial brain stimulation,'' in \emph{Proceedings of the 2019 ACM SIGSAC Conference on Computer and Communications Security}, 2019, pp. 1265--1282.

\bibitem{wang2019neural}
B.~Wang, Y.~Yao, S.~Shan, H.~Li, B.~Viswanath, H.~Zheng, and B.~Y. Zhao, ``Neural cleanse: Identifying and mitigating backdoor attacks in neural networks,'' in \emph{2019 IEEE Symposium on Security and Privacy (SP)}.\hskip 1em plus 0.5em minus 0.4em\relax IEEE, 2019, pp. 707--723.

\bibitem{guo2023universal}
W.~Guo, B.~Tondi, and M.~Barni, ``Universal detection of backdoor attacks via density-based clustering and centroids analysis,'' \emph{IEEE Transactions on Information Forensics and Security}, 2023.

\bibitem{zhu2024neuralsanitizer}
H.~Zhu, Y.~Zhao, S.~Zhang, and K.~Chen, ``Neuralsanitizer: Detecting backdoors in neural networks,'' \emph{IEEE Transactions on Information Forensics and Security}, 2024.

\bibitem{tran2018spectral}
B.~Tran, J.~Li, and A.~Madry, ``Spectral signatures in backdoor attacks,'' \emph{Advances in Neural Information Processing Systems}, vol.~31, 2018.

\bibitem{gao2019strip}
Y.~Gao, C.~Xu, D.~Wang, S.~Chen, D.~C. Ranasinghe, and S.~Nepal, ``Strip: A defence against trojan attacks on deep neural networks,'' in \emph{Proceedings of the 35th Annual Computer Security Applications Conference}, 2019, pp. 113--125.

\bibitem{liu2023detecting}
X.~Liu, M.~Li, H.~Wang, S.~Hu, D.~Ye, H.~Jin, L.~Wu, and C.~Xiao, ``Detecting backdoors during the inference stage based on corruption robustness consistency,'' in \emph{Proceedings of the IEEE/CVF Conference on Computer Vision and Pattern Recognition}, 2023, pp. 16\,363--16\,372.

\bibitem{10646679}
X.~Mo, Y.~Zhang, L.~Y. Zhang, W.~Luo, N.~Sun, S.~Hu, S.~Gao, and Y.~Xiang, ``Robust backdoor detection for deep learning via topological evolution dynamics,'' in \emph{2024 IEEE Symposium on Security and Privacy (SP)}, 2024, pp. 2048--2066.

\bibitem{qi2022revisiting}
X.~Qi, T.~Xie, Y.~Li, S.~Mahloujifar, and P.~Mittal, ``Revisiting the assumption of latent separability for backdoor defenses,'' in \emph{The eleventh international conference on learning representations}, 2022.

\bibitem{9230390}
T.~J.~L. Tan and R.~Shokri, ``Bypassing backdoor detection algorithms in deep learning,'' in \emph{2020 IEEE European Symposium on Security and Privacy (EuroSP)}, 2020, pp. 175--183.

\bibitem{bagdasaryan2021blind}
E.~Bagdasaryan and V.~Shmatikov, ``Blind backdoors in deep learning models,'' in \emph{Usenix Security}, 2021.

\bibitem{peng2024model}
H.~Peng, H.~Qiu, H.~Ma, S.~Wang, A.~Fu, S.~F. Al-Sarawi, D.~Abbott, and Y.~Gao, ``On model outsourcing adaptive attacks to deep learning backdoor defenses,'' \emph{IEEE Transactions on Information Forensics and Security}, 2024.

\bibitem{xie2020dba}
C.~Xie, K.~Huang, P.~Y. Chen, and B.~Li, ``Dba: Distributed backdoor attacks against federated learning,'' in \emph{8th International Conference on Learning Representations, ICLR 2020}, 2020.

\bibitem{xue2020one}
M.~Xue, C.~He, J.~Wang, and W.~Liu, ``One-to-n \& n-to-one: Two advanced backdoor attacks against deep learning models,'' \emph{IEEE Transactions on Dependable and Secure Computing}, vol.~19, no.~3, pp. 1562--1578, 2020.

\bibitem{xiao2022multitarget}
Y.~Xiao, L.~Cong, Z.~Mingwen, W.~Yajie, L.~Xinrui, S.~Shuxiao, M.~Yuexuan, and Z.~Jun, ``A multitarget backdooring attack on deep neural networks with random location trigger,'' \emph{International Journal of Intelligent Systems}, vol.~37, no.~3, pp. 2567--2583, 2022.

\bibitem{hayase2021spectre}
J.~Hayase, W.~Kong, R.~Somani, and S.~Oh, ``Spectre: Defending against backdoor attacks using robust statistics,'' in \emph{International Conference on Machine Learning}.\hskip 1em plus 0.5em minus 0.4em\relax PMLR, 2021, pp. 4129--4139.

\bibitem{dumford2020backdooring}
J.~Dumford and W.~Scheirer, ``Backdooring convolutional neural networks via targeted weight perturbations,'' in \emph{2020 IEEE International Joint Conference on Biometrics (IJCB)}.\hskip 1em plus 0.5em minus 0.4em\relax IEEE, 2020, pp. 1--9.

\bibitem{tang2020embarrassingly}
R.~Tang, M.~Du, N.~Liu, F.~Yang, and X.~Hu, ``An embarrassingly simple approach for trojan attack in deep neural networks,'' in \emph{Proceedings of the 26th ACM SIGKDD international conference on knowledge discovery \& data mining}, 2020, pp. 218--228.

\bibitem{edraki2021odyssey}
M.~Edraki, N.~Karim, N.~Rahnavard, A.~Mian, and M.~Shah, ``Odyssey: Creation, analysis and detection of trojan models,'' \emph{IEEE Transactions on Information Forensics and Security}, vol.~16, pp. 4521--4533, 2021.

\bibitem{chen2018detecting}
B.~Chen, W.~Carvalho, N.~Baracaldo, H.~Ludwig, B.~Edwards, T.~Lee, I.~Molloy, and B.~Srivastava, ``Detecting backdoor attacks on deep neural networks by activation clustering,'' \emph{arXiv preprint arXiv:1811.03728}, 2018.

\bibitem{huang2022backdoor}
\BIBentryALTinterwordspacing
K.~Huang, Y.~Li, B.~Wu, Z.~Qin, and K.~Ren, ``Backdoor defense via decoupling the training process,'' in \emph{International Conference on Learning Representations}, 2022. [Online]. Available: \url{https://openreview.net/forum?id=TySnJ-0RdKI}
\BIBentrySTDinterwordspacing

\bibitem{10.1007/978-3-031-33377-4_33}
N.~M. Jebreel, J.~Domingo-Ferrer, and Y.~Li, ``Defending against backdoor attacks by layer-wise feature analysis,'' in \emph{Advances in Knowledge Discovery and Data Mining}, H.~Kashima, T.~Ide, and W.-C. Peng, Eds.\hskip 1em plus 0.5em minus 0.4em\relax Cham: Springer Nature Switzerland, 2023, pp. 428--440.

\bibitem{doi:10.1073/pnas.0601602103}
\BIBentryALTinterwordspacing
M.~E.~J. Newman, ``Modularity and community structure in networks,'' \emph{Proceedings of the National Academy of Sciences}, vol. 103, no.~23, pp. 8577--8582, 2006. [Online]. Available: \url{https://www.pnas.org/doi/abs/10.1073/pnas.0601602103}
\BIBentrySTDinterwordspacing

\bibitem{Fortunato_2010}
\BIBentryALTinterwordspacing
S.~Fortunato, ``Community detection in graphs,'' \emph{Physics Reports}, vol. 486, no. 3–5, p. 75–174, Feb. 2010. [Online]. Available: \url{http://dx.doi.org/10.1016/j.physrep.2009.11.002}
\BIBentrySTDinterwordspacing

\bibitem{zhao2019pyod}
\BIBentryALTinterwordspacing
Y.~Zhao, Z.~Nasrullah, and Z.~Li, ``Pyod: A python toolbox for scalable outlier detection,'' \emph{Journal of Machine Learning Research}, vol.~20, no.~96, pp. 1--7, 2019. [Online]. Available: \url{http://jmlr.org/papers/v20/19-011.html}
\BIBentrySTDinterwordspacing

\bibitem{CIFAR10}
``The cifar-10 dataset,'' \url{https://www.cs.toronto.edu/~kriz/cifar.html}.

\bibitem{stallkamp2012man}
J.~Stallkamp, M.~Schlipsing, J.~Salmen, and C.~Igel, ``Man vs. computer: Benchmarking machine learning algorithms for traffic sign recognition,'' \emph{Neural networks}, vol.~32, pp. 323--332, 2012.

\bibitem{he2016deep}
K.~He, X.~Zhang, S.~Ren, and J.~Sun, ``Deep residual learning for image recognition,'' in \emph{Proceedings of the IEEE conference on computer vision and pattern recognition}, 2016, pp. 770--778.

\bibitem{deng2009imagenet}
J.~Deng, W.~Dong, R.~Socher, L.-J. Li, K.~Li, and L.~Fei-Fei, ``Imagenet: A large-scale hierarchical image database,'' in \emph{2009 IEEE conference on computer vision and pattern recognition}.\hskip 1em plus 0.5em minus 0.4em\relax Ieee, 2009, pp. 248--255.

\end{thebibliography}
}
\appendices
\section{Bag of Tricks for Adaptive Attacks Data Poisoning}
\label{appendix:Adaptive Attack}

\subsection{\laundry}
The \laundry dataset $\mathcal{D}_\text{l}$ has evolved to adaptively conceal the malicious backdoor effect from $\mathcal{D}_\text{p}$ by introducing a conditional trigger \cite{qi2022revisiting,tang2021demon,lin2020composite,10646679,10191661}. As demonstrated in \cite{peng2024model}, model regularization can also be employed to boost the \laundry by aligning backdoored model's parameters with those of a clean counterpart.
\laundry entails incorporating data that appears similar to the triggers but with correct labels. The goal is to reduce the network's immediate trigger sensitivity by merging trigger features with those of the target class, thus lessening immediate reactivity \cite{10646679}.

\subsection{\slowrelease}
\label{Subsec:slowrelease}

\slowrelease aims to reduce the model's dependence on trigger patterns in individual data instances during training, thereby minimizing the overall visibility of these triggers. This method involves decreasing the trigger prevalence in each training sample while maintaining the full trigger for inference. As discussed in \cite{xie2020dba,xue2020one}, this approach may include fragmenting the trigger into multiple segments to dilute its immediate influence, or gradually ``poisoning'' the model throughout the training process. When the cumulative impact of these segments—termed aggregate toxicity—reaches a certain threshold, it results in an increase in the ASR while preserving the Clean ACC.

\subsection{\targetmapping}
\label{Subsec:targetMapping}

During the poisoning process, the attacker manipulates source-to-target class mappings to control the backdoored model's behavior. This can involve many-to-many mapping, where each true label is assigned a different target label, or many-to-one mapping, where all true labels of triggered data are changed to a fixed target label \cite{edraki2021odyssey}. A backdoored model can switch between malicious tasks based on specific backdoor features within an input \cite{bagdasaryan2021blind}. These features may be the trigger attribute alone (\textit{e.g.}, location \cite{xiao2022multitarget} or intensity \cite{xue2020one}), or a combination of trigger attribute and source class \cite{bagdasaryan2021blind}.

\subsection{\laundry and \slowrelease}

When combining \laundry and \slowrelease techniques, as seen in Adap-Blend and Adap-Patch backdoor attacks \cite{qi2022revisiting}, both the poisoned dataset $\mathcal{D}_\text{p}$ and the \laundry dataset $\mathcal{D}_\text{l}$ utilize the same modification function $A_\text{trig}$ (Eq.~\eqref{Eq:amod}). This results in the following definitions:
\begin{IEEEeqnarray}{rCl} 
\mathcal{D}_\text{p} &=& \{((A_\text{trig}(x, \beta_\text{t}), c_\text{target}) \mid (x,y) \in \mathcal{D}\},  \label{Eq:backdoor dataset combined} \\ 
\mathcal{D}_\text{l} &=& \{((A_\text{trig}(x, \beta_\text{t}), y) \mid (x,y) \in \mathcal{D}\}. 
\label{Eq:laundry dataset combined}
\end{IEEEeqnarray}

\subsection{\slowrelease and \targetmapping}

Combining \slowrelease with two types of \targetmapping, we incorporate \targetmapping into the training process. Using the modified trigger (Eq.~\eqref{Eq:amod}) in Eq.~\eqref{eq:ss_targetmapping} and Eq.~\eqref{eq:ssta_targetmapping}, we define the dataset \(\mathcal{D}_\text{p}\) as follows:
\begin{IEEEeqnarray}{rCl}
\mathcal{D}_\text{p} &=& \big\{(A_\text{trig}(x, \beta_\text{t}), T(y, g(\beta_\text{t}))) \mid (x,y) \in \mathcal{D}, \nonumber \\
&& \quad y \in \mathcal{S}, \beta_\text{t} \in \mathcal{R}_\text{t}\big\},
\end{IEEEeqnarray}
or if considering source-specific only:
\begin{IEEEeqnarray}{l}
\hspace{-2.5em} \mathcal{D}_\text{p} = \big\{(A_\text{trig}(x, \beta_\text{t}), T(y)) \mid (x,y) \in \mathcal{D}, \nonumber \\
\hspace{0em} \quad y \in \mathcal{S}, \beta_\text{t} \in \mathcal{R}_\text{t}\big\}.
\end{IEEEeqnarray}
\subsection{\laundry and \targetmapping}

Combining \laundry with two types of \targetmapping, we randomly select two victim classes, two target classes, and two \laundry classes. The poisoning dataset comprises samples from victim classes, mixed with the trigger and labeled as the target class. The \laundry data, randomly selected from non-victim classes, are mixed with the trigger but retain their original labels.
Integrating Eq.~\eqref{Eq:laundry dataset} with Eq.~\eqref{eq:ss_targetmapping} and Eq.~\eqref{eq:ssta_targetmapping}, we define:
\begin{IEEEeqnarray}{rCl} 
\mathcal{D}_\text{p} &=& \big\{\big(A_\text{trig}(x, \beta), T(y, \beta)\big) \mid (x,y) \in \mathcal{D}, \big. \nonumber \\
&& \big. y \in \mathcal{S} \cup \{\emptyset\}, \beta \in \mathcal{R} \cup \{\emptyset\}\big\},\\
\mathcal{D}_\text{l} &=& \{(A_\text{trig}(x), y) \mid  (x,y) \in \mathcal{D}  \label{Eq:laundry dataset SSTA} \}.
\end{IEEEeqnarray}
For situations that concentrate solely on the source-specific:
\begin{IEEEeqnarray}{rCl}
\mathcal{D}_\text{p} &=& \big\{(A_\text{trig}(x), T(y)) \mid (x,y) \in \mathcal{D}, y \in \mathcal{S} \big\},\\
\mathcal{D}_\text{l} &=& \{(A_\text{trig}(x), y) \mid  (x,y) \in \mathcal{D}  \label{Eq:laundry dataset SS} \}.
\end{IEEEeqnarray}

\subsection{\laundry, \slowrelease and \targetmapping}

We modify the poisoning datasets $\mathcal{D}_\text{p}$ (Eq.~\eqref{eq:ssta_targetmapping}) and \laundry dataset $\mathcal{D}_\text{l}$ (Eq.~\eqref{Eq:laundry dataset}) using the altered trigger defined in Eq.~\eqref{Eq:amod}. For the SS\&TA+L+SR approach, the poisoned dataset $\mathcal{D}_\text{p}$ and the \laundry dataset $\mathcal{D}_\text{l}$ are defined as:
\begin{IEEEeqnarray}{rCl} 
\mathcal{D}_\text{p} &=& \big\{ (A_\text{trig}(x, \beta_\text{t}), T(y, g(\beta_\text{t}))) \mid (x,y) \in \mathcal{D}, 
\nonumber \\
&& y \in \mathcal{S}, \beta_\text{t} \in \mathcal{R}_\text{t}\big\}, \label{eq:SSandTA_L_SR_Dp}\\
\mathcal{D}_\text{l} &=& \big\{ (A_\text{trig}(x, \beta_\text{t}), y) \mid (x,y) \in \mathcal{D}, \beta_\text{t} \in \mathcal{R}_\text{t}\big\}. \label{eq:SSandTA_L_SR_Dl}
\end{IEEEeqnarray}
For scenarios focusing solely on source-specific manipulation (SS+L+SR), the poisoned dataset and the \laundry dataset are constructed as:
\begin{IEEEeqnarray}{rCl} 
\mathcal{D}_\text{p} &=& \big\{ (A_\text{trig}(x, \beta_\text{t}), T(\beta)) \mid (x,y) \in \mathcal{D},  \beta_\text{t} \in \mathcal{R}_\text{t}, \beta \in \mathcal{R}\big\}, \IEEEeqnarraynumspace \label{eq:SS_L_SR_Dp}  \\
\mathcal{D}_\text{l} &=& \big\{ (A_\text{trig}(x, \beta_\text{t}), y) \mid (x,y) \in \mathcal{D},  \beta_\text{t} \in \mathcal{R}_\text{t}\big\}.  \label{eq:SS_L_SR_Dl}
\end{IEEEeqnarray}

\begin{figure}
\centering
\begin{tabular}{cc}
\includegraphics[width=0.25\linewidth, trim={0 0 0 150pt}, clip]{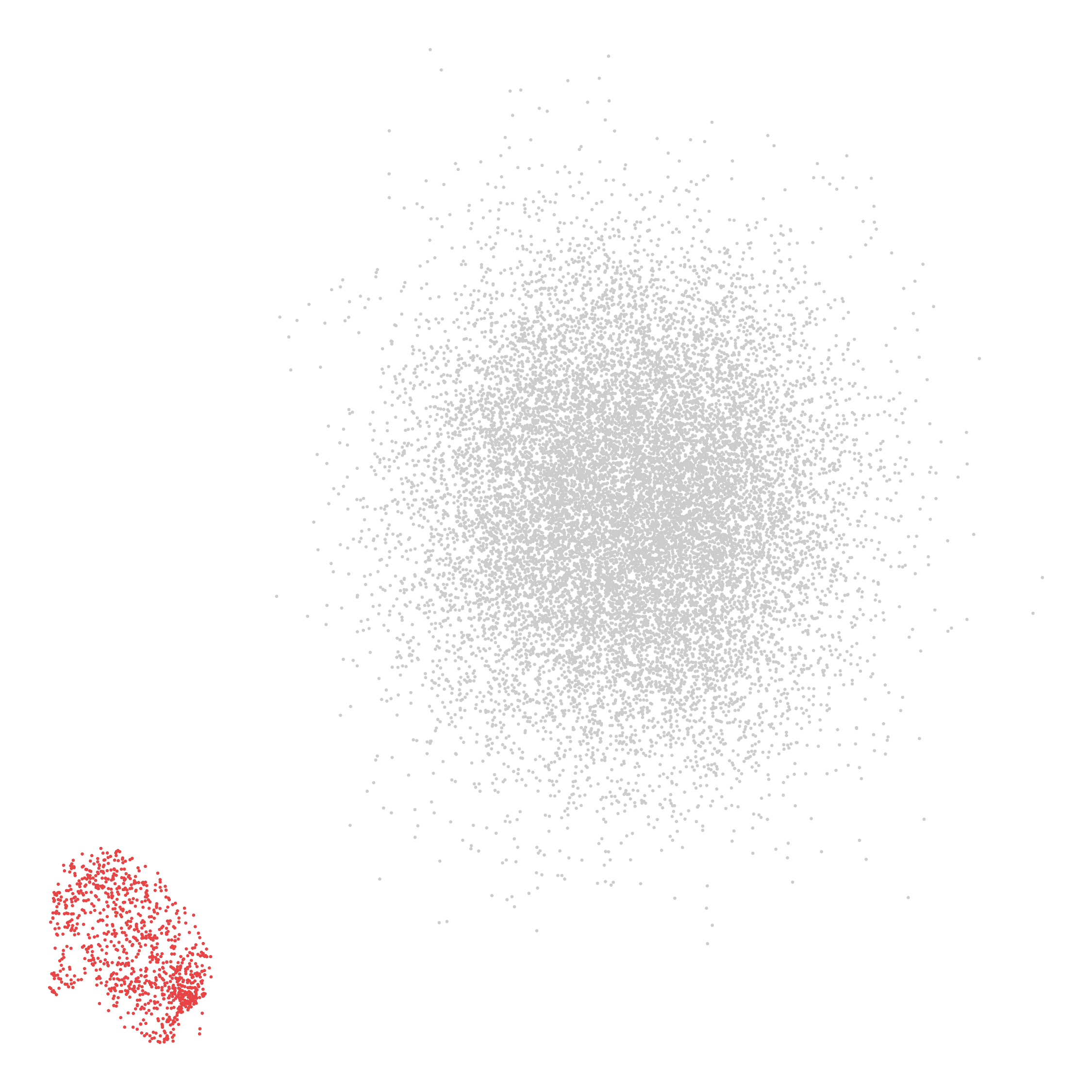} &
\includegraphics[width=0.25\linewidth, trim={0 0 0 150pt}, clip]{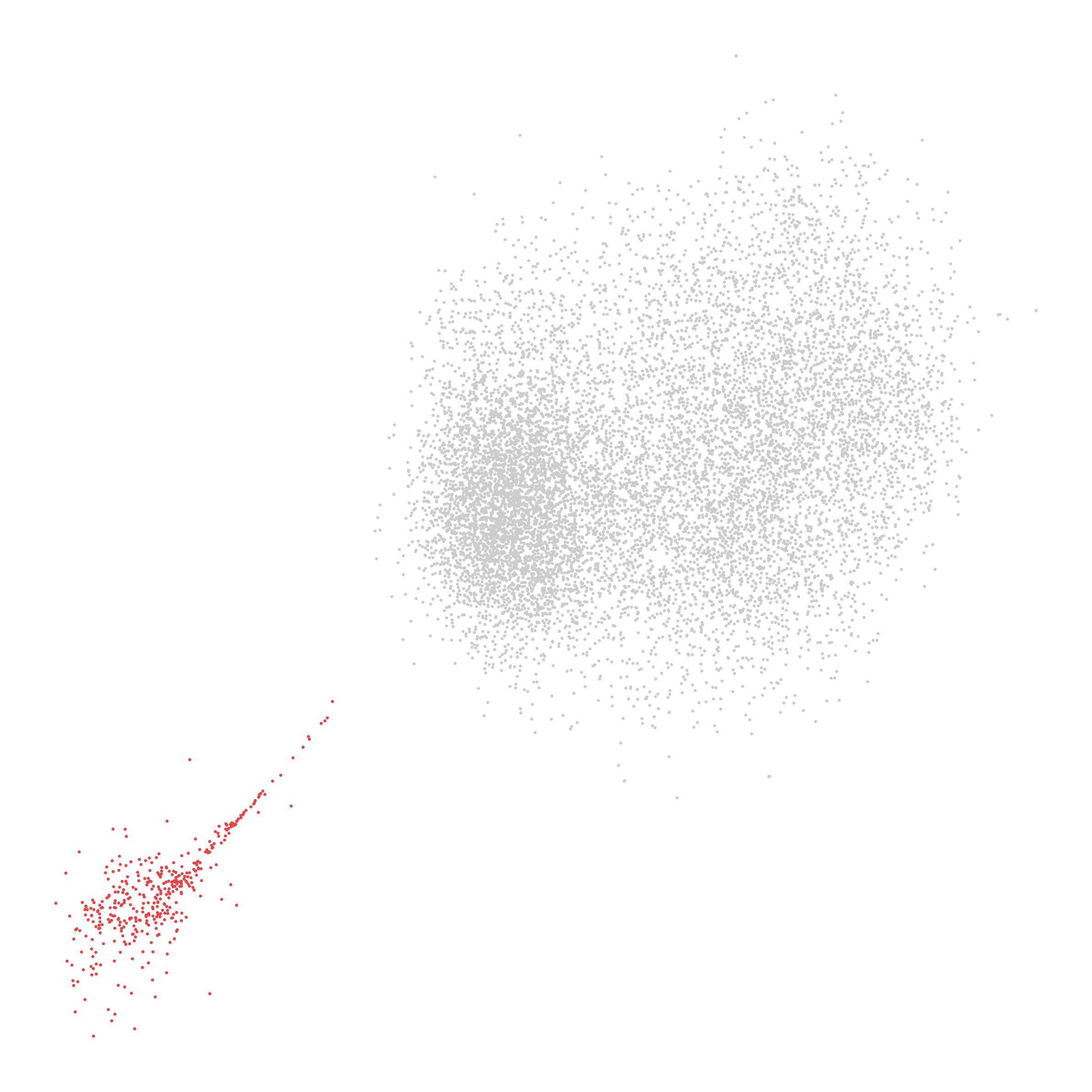} \\
(a) BadNets & (b) SSDT 
\end{tabular}
\caption{T-SNE visualization of \ourdefense feature vectors for BadNets and SSDT attacks on CIFAR-10. Red points represent malicious samples and gray points represent clean samples.}
\label{fig:tsne_plots_non_adaptive_attack}
\end{figure}

\section{Adaptive Weighting for Dynamic Target Classes Under Subtle Perturbations}
\label{appendix:Subtle Perturbations on Noise Quantification and Adaptive Weighting}

We evaluate our adaptive weighting methods on SS\&TA+L+SR samples with subtle perturbations. Fig.~\ref{fig:ssta+l+srweightingcompare} compares topological feature vectors across network layers, with and without our weighting method, for two dynamic target classes.
The non-weighted scenario (Fig.~\ref{fig:ssta+l+srweightingcompare}a and \ref{fig:ssta+l+srweightingcompare}c) shows significant overlap between clean and malicious sample features, especially in early layers, due to SS\&TA+L+SR's subtle perturbations being masked by inherent target class variations.
Our modularity-based weighting method (Fig.~\ref{fig:ssta+l+srweightingcompare}b and \ref{fig:ssta+l+srweightingcompare}d) addresses this by adaptively adjusting layer importance. It identifies layers with clearer distinctions between clean and malicious samples, assigning higher weights to amplify discriminative features. The visualization demonstrates improved separation between clean and malicious samples for both dynamic target classes, particularly in later layers where our weighting scheme has the greatest impact.

\section{Robustness Against More Existing Backdoor Attacks}
\label{appendix:Non-Adaptive Attack}

We evaluate the robustness of \ourdefense against two backdoor attacks: BadNets and SSDT. BadNets \cite{gu2017badnets}, the most common backdoor attack, uses a gray square on the bottom right as the trigger with a 0.01 poison rate. For Source-Specific Dynamic Trigger (SSDT) \cite{10646679}, which is an attack that applies \laundry and also uses dynamic triggers, poison and clean samples consistently form two separate clusters in the topological space. SSDT employs source-specific dynamic triggers, activating only for victim class samples to reduce impact on non-victim classes.
For SSDT, we replicated the original experimental settings using the authors' source code on CIFAR-10. \ourdefense achieved an AUROC of 1 for both BadNets and SSDT, indicating perfect separation of clean and poisoned samples. Fig.~\ref{fig:tsne_plots_non_adaptive_attack} visualizes this separation through T-SNE plots.

\end{document}